\documentclass[num-refs]{wiley-article}

\usepackage{amsmath}
\usepackage{amssymb}
\usepackage[binary-units=true]{siunitx}
\usepackage{url}
\usepackage{subfig,subfloat}
\usepackage{booktabs}
\usepackage{microtype}
\usepackage{multicol}
\usepackage{dcolumn}
\usepackage{pgfplots}
\usepackage{multirow}
\usepackage{float}
\newcommand\uc[1]{\MakeUppercase{#1}}

\usepackage{listings}
\lstdefinestyle{customcpp}{%
  belowcaptionskip=1\baselineskip,
  breaklines=true,
  xleftmargin=\parindent,
  language=C++,
  showstringspaces=false,
  basicstyle=\linespread{0.4}\small\ttfamily,
  keywordstyle=\bfseries\color{green!40!black},
  numberstyle=\tiny,
  commentstyle=\itshape\color{purple!40!black},
  identifierstyle=\bfseries\color{black},
  stringstyle=\color{red},
  emph={int,char,double,float,unsigned},
  emphstyle=\color{blue},
  morekeywords={uint64_t,uint32_t,__m256i,__m128i,UINT64_C},
}
\restylefloat{table}
\date{}
\newcolumntype{g}{D{.}{.}{-1}}
\newcommand{\mc}[1]{\multicolumn{1}{c}{\emph{#1}}}
\newcommand{\dash}{{-{}-{}-{}-}}
\usepackage{tikz}
\usepackage{tikzscale}
\newcommand*{\instruction}[1]{\texttt{#1}}
 \usepackage{color,soul}

\newcommand*{\fourgramt}[8]{\texttt{#1\,#2\;#3\,#4\;#5\,#6\;#7\,#8}}
\newcommand*{\wordgramt}[4]{\texttt{#1\,#2\,#3\,#4}}
\newcommand*{\codepoint}[1]{\texttt{U+\MakeUppercase{#1}}}
\newcommand*{\codepointrange}[2]{\texttt{U+\MakeUppercase{#1}}--\texttt{U+\MakeUppercase{#2}}}
\newcommand*{\hexrange}[2]{\texttt{0x\MakeUppercase{#1}}--\texttt{0x\MakeUppercase{#2}}}

\definecolor{bblue}{HTML}{4F81BD}
\definecolor{rred}{HTML}{C0504D}
\definecolor{ggreen}{HTML}{9BBB59}
\definecolor{ppurple}{HTML}{9F4C7C}

\DeclareMathOperator{\width}{width}
\DeclareMathOperator{\ctz}{ctz}
\DeclareMathOperator{\pdep}{pdep}
\DeclareMathOperator{\pext}{pext}
\DeclareMathOperator{\compress}{compress}
\DeclareMathOperator{\popcount}{popcount}
\newcommand*{\restartrowcolors}{%
  \ifhmode\unskip\fi
  \vadjust{%
    \global\rownum=0 %
  }%
}
\title{Transcoding Unicode Characters with AVX-512 Instructions}

\runningauthor{Robert Clausecker and Daniel Lemire}

\author[1\authfn{1}]{Robert Clausecker}
\affil[1]{Zuse Institute Berlin, Germany}
\author[2\authfn{2}]{Daniel Lemire}
\affil[2]{DOT-Lab Research Center, Université du Québec (TELUQ), Montréal, Canada}

\corraddress{Daniel Lemire, DOT-Lab Research Center, Universit\'e du Qu\'ebec (TELUQ), Montreal, Quebec, H2S 3L5, Canada}
\corremail{daniel.lemire@teluq.ca}
\fundinginfo{Natural Sciences and Engineering Research Council of Canada, Grant Number: RGPIN-2017-03910}
\usepackage{xpatch}
\makeatletter
\xpreto\tabular{\global\rownum=\z@}
\makeatother
\begin{document}
\maketitle
\begin{abstract}
Intel includes in its recent processors a powerful set of instructions
capable of processing 512-bit registers with a single instruction (AVX-512).
Some of these instructions have no equivalent in earlier instruction sets.
We leverage these instructions to efficiently transcode strings between the most
common formats: UTF-8 and UTF-16.
With our novel algorithms, we are often twice as fast as the previous best solutions.
For example, we transcode Chinese text from UTF-8 to UTF-16 at more than  
\SI{5}{\gibi\byte\per\second} using fewer 
than 2~CPU instructions per character. 
To ensure reproducibility, we make our software freely available as an open source library. Our library is part of the popular Node.js JavaScript runtime.

\end{abstract}

\section{Introduction}
Computers store strings of text as arrays of bytes. Unicode is a standard for representing text as a sequence of \emph{universal characters} represented by \emph{code points}. Code points are stored as short sequences of bytes according to a given \emph{unicode transformation format} (UTF), the most popular being UTF-8 and UTF-16.
Not all sequence of bytes are valid UTF-8 or UTF-16 strings~\cite{keiser2020validating}.
Validation is required to detect incorrectly encoded or corrupted text before it is processed.

We often need to transcode strings between the two formats. 
For example, a database might store data in UTF-16 and yet the
programmer might need to produce UTF-8 strings for a web site.
Thankfully, transcoding is relatively efficient.
Conventional transcoders often achieve a throughput between \SIrange{0.5}{1.5}{\gibi\byte\per\second} on commodity processors~\cite{lemire2022transcoding} (cf.~\S~\ref{sec:experiments}).\footnote{The speed is measured by taking the size of the input and dividing by the time elapsed.} Yet this falls far below the sequential-read speed of a fast disk (e.g., \SI{5}{\gibi\byte\per\second}) or the throughput of a fast network connection.

IBM mainframes based on z/Architecture provide special-purposes instructions named ``\uc{convert utf-8 to utf-16}'' and ``\uc{convert utf-16 to utf-8}'' for translation between the two encodings~\cite{zpop}.
By virtue of being implemented in hardware, these exceed \SI{10}{\gibi\byte\per\second} processing speed for typical inputs.
While commodity processors currently lack such dedicated instructions, they can benefit from single-instruction-multiple-data (SIMD) instructions.
Unlike conventional instructions which operate on a single machine word (e.\,g. 64~bits), these SIMD instructions operate on larger registers (128~bits, 256~bits, \ldots) representing vectors of numbers.
A single SIMD instruction may add eight pairs of 16-bit words at once.
We can transcode gigabytes of text per second~\cite{lemire2022transcoding} by a deliberate use  of conventional SIMD instructions (e.g., ARM NEON, SSE, AVX2).

In recent years, Intel introduced new SIMD instruction sets operating over registers as wide as 512~bits. 
If Intel had merely doubled the width of the registers, there would be little need for further work on our part.
However, our experience suggests that to fully benefit from AVX-512 instructions, we need to use adapted algorithms~\cite{mula2020base64}.
Indeed, while AVX-512 instructions benefit from wider registers, Intel has also added many more instructions than what is typically found in SIMD instruction sets.
There is also a slightly different model: AVX-512 instruction may consume or generate masks in \emph{mask} registers which have no equivalent in prior commodity instruction sets. In AVX-512, a mask is conceptually an array of 8, 16, 32, or~64~bits corresponding to vectors of 8, 16, 32, or~64~elements.

We present novel transcoding functions using AVX-512 instructions. On average, we are roughly twice as fast as the previous fastest functions~\cite{lemire2022transcoding} on commodity processors.


\begin{figure}[tbh!]
\centering
\subfloat[\label{16to8:correspondence}Bit-by-bit correspondence between UTF-16 and UTF-8 encodings in the four possible cases. The bits are named~\texttt A\ to~\texttt W\ starting at the least significant bits with $\mathtt{0vuts}=\mathtt{WVUTS}-1$.]{
\begin{tabular}{rcc}
\toprule
\mc{case}&\mc{UTF-16}&\mc{UTF-8}\\
\midrule
ASCII&\wordgramt{0000}{0000}{0GFE}{DCBA}&\fourgramt\dash\dash\dash\dash\dash\dash{\underline{0}GFE}{DCBA}\\
2-byte&\wordgramt{0000}{0LKJ}{HGFE}{DCBA}&\fourgramt\dash\dash\dash\dash{\underline{110}L}{KJHG}{\underline{10}FE}{DCBA}\\
3-byte&\wordgramt{RQPN}{MLKJ}{HGFE}{DCBA}&\fourgramt\dash\dash{\underline{1110}}{RQPN}{\underline{10}ML}{KJHG}{\underline{10}FE}{DCBA}\\
4-byte&\wordgramt{\underline{1101}}{\underline{10}vu}{tsRQ}{PNML}&\fourgramt{\underline{1111}}{\underline{0}WVU}{\underline{10}TS}{RQPN}{\underline{10}ML}{KJHG}{\underline{10}FE}{DCBA}\\
      &\wordgramt{\underline{1101}}{\underline{11}KJ}{HGFE}{DCBA}&                                                          \\
\bottomrule
\end{tabular}
}\\
\subfloat[Examples of matched code-point values in UTF-32, UTF-16LE and UTF-8. For \codepoint{10000} and \codepoint{10FFFF}, UTF-16 requires a surrogate pair. \label{16to8:example}]{
\setlength\tabcolsep{.375em}
\begin{tabular}{rcc}
\toprule
\mc{codepoint}&\mc{UTF-16}&\mc{UTF-8}\\
\midrule
\codepoint{0}&\wordgramt{0000}{0000}{0000}{0000}&\fourgramt\dash\dash\dash\dash\dash\dash{\underline{0}000}{0000}\\
\codepoint{7F} &\wordgramt{0000}{0000}{0111}{1111}&\fourgramt\dash\dash\dash\dash\dash\dash{\underline{0}111}{1111}\\
\codepoint{80}&\wordgramt{0000}{0000}{1000}{0000}&\fourgramt\dash\dash\dash\dash{\underline{110}0}{0010}{\underline{10}00}{0000}\\
\codepoint{7FF}&\wordgramt{0000}{0111}{1111}{1111}&\fourgramt\dash\dash\dash\dash{\underline{110}1}{1111}{\underline{10}11}{1111}\\
\codepoint{800}&\wordgramt{0000}{1000}{0000}{0000}&\fourgramt\dash\dash{\underline{1110}}{0000}{\underline{10}10}{0000}{\underline{10}00}{0000}\\
\codepoint{FFFF}&\wordgramt{1111}{1111}{1111}{1111}&\fourgramt\dash\dash{\underline{1110}}{1111}{\underline{10}11}{1111}{\underline{10}11}{1111}\\
\codepoint{10000}&\wordgramt{\underline{1101}}{\underline{10}00}{0000}{0000}&\fourgramt{\underline{1111}}{\underline{0}000}{\underline{10}01}{0000}{\underline{10}00}{0000}{\underline{10}00}{0000}\\
      &\wordgramt{\underline{1101}}{\underline{11}00}{0000}{0000}&                                                          \\
\codepoint{10FFFF}&\wordgramt{\underline{1101}}{\underline{10}11}{1111}{1111}&\fourgramt{\underline{1111}}{\underline{0}100}{\underline{10}00}{1111}{\underline{10}11}{1111}{\underline{10}11}{1111}\\
      &\wordgramt{\underline{1101}}{\underline{11}11}{1111}{1111}&                                                          \\
\bottomrule
\end{tabular}
}
\caption{Correspondence between UTF-16 and UTF-8. Format-specific prescribed bits (\emph{tag~bits}) are underlined.  \label{fig:16to8:correspondenceoverall}}
\end{figure}

\section{Unicode and its Encodings}
Unicode is a standard based on the \emph{Universal Character Set}~(UCS).
An extension to~ASCII, UCS~is a character set whose characters (called \emph{universal characters}) have code points numbered from \codepoint{0}\footnote{\texttt{U+} followed by a hexadecimal number is notation for a universal character's code point.} to~\codepoint{10FFFF} (decimal~\num{1114111}).
These code points are organized into 17~\emph{planes} of \num{65536}~characters each, with the first plane~\codepointrange{0000}{ffff} being called the~\emph{Basic Multilingual Plane}~(BMP).
Code points in the range \texttt{0xd800}--\texttt{0xdfff} are reserved for \emph{surrogates} used in the UTF-16 encoding and do not represent universal characters.

Unlike simpler character sets like ASCII, universal characters are seldomly stored directly as integers, as such a storage format is wasteful and incompatible to existing byte-oriented environments.
Instead, several \emph{Unicode Transformation Formats}~(UTF) are employed to store and process universal characters, depending on the use case at hand.

A Unicode Transformation Format transforms each universal character into a sequence of integers, with the size of the integer being dependent on the format.  
Popular Unicode Transformation Formats include:
\begin{description}
\item[UTF-32] representing each universal character as a 32-bit integer.  Mainly used as an internal representation.
\item[UTF-16] representing each universal character as one or two 16-bit integers~\cite{rfc2781}.  All code-point values up to \codepoint{FFFF} are stored as 2-byte integer values directly. Otherwise we use surrogate pairs: two consecutive 2-byte values, each storing 10~bits of the codepoint.  Used by Java, Windows~NT, databases, binary protocols, and others.
\item[UTF-8] representing each universal character as 1--4 bytes~\cite{rfc3629}.  An extension to ASCII, UTF-8 is by far the most popular text encoding on the World Wide Web.
\end{description}
Though our software work covers many cases (from UTF-8 to UTF-16 or UTF-32, from UTF-16 to UTF-8 or UTF-32, and so forth), we study the two most difficult  cases: from UTF-8 to UTF-16 and back. 

Multi-byte words in computers representing numerical
values can be stored in either
little-endian format or big-endian format, depending on whether the first byte is the least significant or the most significant.
Unicode Transformation Formats representing characters in units larger than bytes are subject to endianess.
If the endianess is not known from the context\footnote{Big endian is the prescribed default byte order~\cite{rfc2781}, although it is less common.}, it can be given by adding a LE or BE suffix to the name of the Unicode Transformation Format, giving e.\,g.~UTF-16BE or UTF-32LE\@.
We can reverse the order of the bytes---between big and little endian---at high speed: e.g., using one instruction per 64~bytes. 
For simplicity, we  present our results on UTF-8 and UTF-16LE\@.

\subsection{UTF-16} \label{unicode:utf-16}
When the Universal Character Set was initially defined, it was meant to be a 16-bit character set with UTF-16 being its natural encoding, representing each universal character in one 16-bit word.
It was later realized that \num{65536}~code points are insufficient to represent the writing systems of the world's many cultures, especially when having to account for over \num{50000}~Chinese, Japanese, and Korean ideographs.
UCS~was therefore extended past the Basic Multilingual Plane to code points up to~\codepoint{10FFFF} and UTF-16 retrofitted with a \emph{surrogate} mechanism to permit representation of these newly added characters.

UTF-16 is a versatile Unicode Transformation Format as it permits (absent surrogates) easy processing of text in many popular languages, while not being as memory-hungry as UTF-32.
It is widely used in databases and binary file formats and is the preferred internal text representation on Windows~NT\@.
Nevertheless, with the advent and growing popularity of universal characters outside of the~Basic Multilingual Plane, UTF-16 has been steadily declining in use.

Despite big-endian byte order being prescribed for UTF-16, the little-endian variant UTF-16LE is more commonly encountered under the influence of x86's little-endian orientation.
A common convention to deal with this ambiguity is to prefix UTF-16 encoded documents with the \emph{byte order mark} (BOM)~\codepoint{FEFF}.\footnote{\codepoint{FEFF}~only has this function as the first character of a document.  In other positions, it should be treated as an ordinary universal character and must not be stripped or altered.}
Its byte-swapped counterpart~\codepoint{FFFE} is a reserved ``uncharacter'' and should not occur in Unicode text.
If a UTF-16 encoded document begins with~\codepoint{FFFE}, it can thus be assumed to be in wrong byte order, permitting automatic byte-order detection in many situations.
Our algorithms do not make use of this convention and strictly assume UTF-16LE throughout.
A~BOM is neither generated, nor checked for, nor stripped.

As illustrated in the ``UTF-16'' column of Fig.~\ref{fig:16to8:correspondenceoverall}, code points in the Basic Multilingual Plane are represented as themselves.
Code points outside of this plane have~\texttt{0x10000} subtracted from them (the~\emph{surrogate plane shift}), yielding a 20-bit number.
This number is split into two 10-bit halves.
The high half is tagged with~\texttt{0xd800}, yielding a~\emph{high surrogate}.
Likewise, the low half is tagged with~\texttt{0xdc00}, yielding a~\emph{low surrogate}.
The character is then encoded by giving its high surrogate, directly followed by its low surrogate.
It is for this purpose that code points in the range~\texttt{0xd800}--\texttt{0xdfff} do not represent universal characters.

Decoding UTF-16 is a matter of joining the bits of surrogate pairs, leaving Basic-Multilingual-Plane characters unchanged.
Care must be taken to validate that each high surrogate is succeeded by a low surrogate and vice versa.
With this sequencing requirement ensured, all UTF-16 sequences are valid and have a 1:1~mapping to code points.

\begin{table}[tb!]
\centering
\caption{Types of UTF-8 bytes with tag bits underlined.} \label{table:tagbits}
\begin{tabular}{rcc}
\toprule
\mc{type}&\mc{range}&\mc{pattern}\\
\midrule
ASCII lead byte&\texttt{0x00}--\texttt{0x7f}&\texttt{\underline 0XXX\,XXXX}\\
continuation byte&\texttt{0x80}--\texttt{0xbf}&\texttt{\underline{10}XX\,XXXX}\\
2-byte lead byte&\texttt{0xc2}--\texttt{0xdf}&\texttt{\underline{110}X\,XXXX}\\
3-byte lead byte&\texttt{0xe0}--\texttt{0xef}&\texttt{\underline{1110}\,XXXX}\\
4-byte lead byte&\texttt{0xf0}--\texttt{0xf4}&\texttt{\underline{1111\,0}XXX}\\
\bottomrule
\end{tabular}
\end{table}

\subsection{UTF-8}
The most popular Unicode Transformation Format is UTF-8, representing each universal character as a sequence of 1--4 bytes.
Replacing the earlier UTF-1, the format was designed to be backwards-compatible to ASCII while also being safe for use in UNIX file names, and comes with many other desirable features.
Under many circumstances, UTF-8 text can be processed as if it was a conventional ASCII-based 8-bit encoding like those of the ISO-8859 family.
This includes common applications like concatenation, substring search, field-splitting (with ASCII characters or UTF-8 strings for separators), and collation, rendering it the most popular UTF\@.

UTF-8 can be seen as an extension to ASCII, where each ASCII character (\codepointrange{00}{7F}) is represented as itself with other characters being represented by sequences of bytes in the range~\texttt{0x80}--\texttt{0xf4} (cf.~Table~\ref{table:tagbits}).
Such sequences start with a \emph{lead byte} (\texttt{0xc2}\footnote{\texttt{0xc0} and \texttt{0xc1} would introduce 2-byte sequences corresponding to ASCII characters, which are encoded as single bytes instead.}--\texttt{0xf4}) indicating the length of the sequence in its~\emph{tag bits}, followed by 1--3~\emph{continuation bytes} (\texttt{0x80}--\texttt{0xbf}), making the encoding stateless, and self-synchronizing.

The details are summarized in the ``UTF-8''~column of Fig.~\ref{fig:16to8:correspondenceoverall}:
The bits of the code point are numbered \texttt A--\texttt{W} starting at the least significant bit.
For each of the four possible cases (the ASCII/1-byte case, the 2-byte case, the 3-byte case, and the 4-byte case\footnote{the 1--3-byte cases represent Basic-Multilingual-Plane characters, the 4-byte case corresponds to characters represented as surrogate pairs in UTF-16.}), the bits of the code point are copied into the lead and continuation bytes as indicated in the figure.
Tag bits are applied (underlined in Fig.~\ref{fig:16to8:correspondenceoverall}) to distinguish ASCII, lead, and continuation bytes.

For many universal characters, more than one encoding seems to be possible according to the figure.
However, only the shortest possible encoding for each character is permitted to ensure uniqueness of the encoding.
While 4-byte sequences could encode code points in excess of~\codepoint{10FFFF}, such sequences are not legal either.
The bytes~\texttt{0xc0}, \texttt{0xc1}, and~\texttt{0xf5}--\texttt{0xff} are thus not used by UTF-8.

Decoding UTF-8 begins by looking at the tag bits to tell the start and length of each sequence.
Then, the code point is assembled from the payload of these bytes.
A critical part in decoding UTF-8 is validation, especially against overly-long sequences and illegal code points (surrogates, code points greater than~\texttt{10FFFF}).
In the algorithm presented in~\S~\ref{sec:8to16} we demonstrate how decoding UTF-8 with comprehensive validation and then reencoding it into UTF-16 can be implemented efficiently, leveraging AVX-512 instructions.

\begin{table}
\centering
\caption{Summary of notation\label{table:notation:logic}} 
\begin{tabular}{rl}
   \toprule
  \mc{expression} & \mc{description} \\\midrule
$\lnot a$&bitwise complement of~$a$\\
$\ctz(a)$&number of trailing zeroes in~$a$\\
$\width(a)$&number of bits needed to represent~$a$\\
$\popcount(a)$&number of bits set in~$a$\\
$\pext(a,b)$&the bits given in~$a$ extracted from~$b$\\
$\pdep(a,b)$&$b$~deposited into the bits given in~$a$\\
$\compress(m, v)$&vector~$v$ compressed by mask~$m$\\
$a+b$&sum of $a$ and~$b$\\
$a\ll b$&$a$~logically shifted to the left by $b$~places\\
$a\gg b$&$a$~logically shifted to the right by $b$~places\\
$a=b$&mask indicating elements of~$a$ equal to those of~$b$\\
$a\land b$&bitwise and of $a$ and~$b$\\
$a\lor b$&bitwise or of $a$ and~$b$\\
$a\oplus b$&bitwise exclusive-or of $a$ and~$b$\\
$a\mathrel ?b:c$&ternary operator; equal to $a\land b\lor\lnot a\land c$\\
\bottomrule
\end{tabular}
\end{table}

\section{Related Work}
\label{sec:relatedwork}

There are relatively few academic publications on Unicode string processing using SIMD instructions. 
Cameron~\cite{cameron2008case} proposed a  UTF-8 to UTF-16 transcoder using SIMD instruction using \emph{bit streams}. A bit stream is a transposition on the character inputs. For example, from 128~bytes of data, we produce eight 128-bit registers with the first register containing the most significant bits of each input byte, and the last register containing the least significant bit of each input byte. The transcoding from UTF-8 to UTF-16 is done in this bit stream form with a final phase where unused bytes are removed.  Inoue et al.~\cite{Inoue2008} presented a limited UTF-8 to UTF-16 transcoder which lacked validation and could not handle 4-byte UTF-8 characters. They rely on a \SI{105}{\kibi\byte}~lookup table.

Lemire and Mu{\l}a~\cite{lemire2022transcoding} presented a generic approach that does full UTF-8 to UTF-16 and UTF-16 to UTF-8 transcoding, with validation. Their UTF-8 to UTF-16 transcoding function is similar in principle to the strategy used by Inoue et al.~\cite{Inoue2008} in that they 
rely on the presence of instructions to quickly permute bytes within a register in an arbitrary manner, based on a lookup table.
The  accelerated  UTF-8 to UTF-16 transcoding algorithm  processes up to 12~input UTF-8 bytes at a time. Given the input bytes, it finds beginning of each character, forming a 12-bit word which is used as a key in a 1024-entry table. Each entry in the table contains the number of UTF-8 bytes to consume and an index into another table where we find \emph{shuffle masks}. The tables use about~\SI{11}{\kibi\byte}.
The shuffle masks are applied to the 12~input bytes to form a vector register that can be transformed efficiently. This 12-byte routine works within 
64-byte blocks. The 64-byte blocks are validated using a fast technique~\cite{keiser2020validating}.
Their UTF-16 to UTF-8 algorithm iteratively reads a block of input bytes in a SIMD register. 
Depending of the values of 16-bit words, the algorithm uses one of several paths. E.g., if all 16-bit words are in the range~\codepointrange{0000}{07ff}, the 16-bit words are converted to 32-bit words to ultimately produce 1-byte, 2-byte or 3-byte characters. A series of lookup tables allow the efficient permutations, using a total of~\SI{8.5}{\kibi\byte}.

Gatilov~\cite{stgatilov} produced one of the best and most complete software library for Unicode transcoding~(utf8lut). It is similar in spirit to the work of Lemire and Mu{\l}a~\cite{lemire2022transcoding}, but utf8lut requires larger tables: \SI{2}{\mebi\byte}~for the UTF-8 to UTF-16 transcoder and \SI{16}{\kibi\byte}~for the UTF-16 to UTF-8 transcoder.

Unlike this prior work, our proposals do not require lookup tables. This is possible through the use of novel \emph{compression instructions} introduced with AVX-512VBMI2 (see~\S~\ref{notation:special}, Tbl.~\ref{table:ref:simdinstructions}), allowing us to move bytes to the right places within registers entirely in hardware, without in-memory tables.

\section{Notational Conventions} \label{notation}
In the algorithms described below, all logical symbols refer to bitwise logic.
Comparisons are performed between corresponding elements of vectors, yielding a bit mask of those elements for which the comparison holds.
All arithmetic operations, shifts, and comparisons are performed on unsigned numbers.
The width of the number depends on the vector used.

As a general convention, scalars, vectors of bytes, and masks derived from them are indicated with lowercase letters.
Vectors of 16-~or 32-bit words are indicated with uppercase letters.\footnote{The convention attempts to underline that byte vectors correspond to UTF-8 whereas word vectors correspond to UTF-16.}
The symbol~$n$ is number of bytes in a vector; for AVX-512 it is
$n=64$.
This convention permits us to explain the algorithms in terms of AVX-512 instructions while giving generic formul\ae\ potentially applicable to other future instruction sets.

The operator precedence follows C~precedence rules with
\begin{equation*}
a+b\ll c=d\land e\lor f
\end{equation*}
being parsed as
\begin{equation*}
((((a+b)\ll c)=d)\land e)\lor f.
\end{equation*}
Table~\ref{table:notation:logic} gives a list of symbols used in decreasing order of precedence.

\subsection{Mask Operations}

Masks are conceptually arrays of bits---containing between 8~and 64~bits---meant 
to be used in conjunction with vectors having the same number of elements. 
For example, \emph{byte masks} (noted $m_1, m_{234},\ldots$) may contain 64~bits if  they correspond
to vectors of 64~bytes. We also have \emph{word masks} (e.g., $M_3$) containing 16~bits when they corresponding to 512-bit vectors of 32-bit values. See Appendices~\ref{appendix:utf8utf16} and~\ref{appendix:utf16utf8} for detailed lists of our masks and other variables.
We operate on masks as if they were unsigned integer values: $m_{+3}=m_4\ll3$~means that the whole mask~$m_4$ is shifted to the left by three places to give~$m_{+3}$; the 64~individual mask bits are always either 0 or~1.  The logical operations 
\emph{or}~($\lor$),  \emph{and}~($\land$) and \emph{not}~($\lnot$) are applied bitwise. We have that~$m=0$ sets all bits to zero whereas~$m=\lnot0$ sets all bits to one.

In practice, the processor has several instructions dedicated to AVX-512 mask registers (e.\,g., \texttt{kshiftrd}, \texttt{kandq}, \texttt{korb}).  Mask registers can be converted back and forth to general-purpose registers as needed---with the caveat that the conversion from mask registers to general-purpose registers may have a high latency (e.\,g., 3~cycles).

\subsection{Vector Operations}
When operating on vectors, equations have to be read as ``SIMD formul\ae'' applying element-by-element.
For example, we write
\begin{equation*}
w=m\mathrel?a+b:c
\end{equation*}
to mean ``each element of~$w$ is set to the sum of the corresponding elements in $a$ and~$b$ if the corresponding bit is set in~$m$ or to~$c$ otherwise.''
With an explicit index~$i=0\ldots n-1$, the previous expression could be written as
\begin{equation*}
w[i]=m\land1\ll i\mathrel ?a[i]+b[i]:c[i]\quad\mbox{for }i=0,1,\ldots,n-1.
\end{equation*}
We believe that the presentation as ``SIMD formul\ae'' is easier to understand and prefer it where possible.
Explicit indices are only used when permutations are involved.
For example, we write
\begin{equation*}
w[i]=v\bigl[p[i]\bigr]
\end{equation*}
to mean ``$w$~is $v$~permuted by the index vector~$p$.''

Conversions from one element size to another are not explicitly written out; watch the letter case of the variables used to see when this happens.
All such conversions are zero-extensions or truncations.

\begin{remark}
The conventional binary notation presents the least significant bits last. When working with masks and vectors, these least significant bits correspond to the first elements of the vectors. This discrepancy in the order is a source of confusion, but it is difficult to avoid. Intel intrinsic functions reflect this confusion by providing two sets of functions to create new vectors: \texttt{\_mm512\_set\_*}
and \texttt{\_mm512\_rset\_*} depending on the prefered order~\cite{intelintrinsic}.
\end{remark}

\subsection{Special Functions} \label{notation:special}
We use several special bit-manipulation functions corresponding to instructions available on contemporary x86 computers:
\begin{description}
\item[ctz] The \emph{count trailing zeroes} operation~$\ctz(a)$ counts the number of trailing (least significant) zero bits in~$a$, i.\,e.~how often $a$~can be divided by~$2$ until leaving an odd number.
  It corresponds to the \texttt{bsf}/\texttt{tzcnt} instructions of the x86~instruction set.
  Our algorithms never invoke~$\ctz(0)$.
\item[width]
  The \emph{bit width} operation~$\width(a)$ counts the number of bits needed to represent~$a$.
  It is
\begin{equation}
\width(a)=(a\ne0)\mathrel?\lfloor\log_2a\rfloor+1:0.
\end{equation}
  This operation is efficiently implemented on many architectures through the \emph{count leading zeroes} operation (x86~instruction \texttt{bsr}/\texttt{lzcnt}).
  Our algorithms never invoke~$\width(0)$.
\item[popcount] The \emph{population count} operation~$\popcount(a)$ computes the number of bits set in~$a$.
  This can also be understood as the sum of the bits of~$a$.
  It corresponds to the~\texttt{popcnt} instruction of the x86~instruction set.
\item[pext] The \emph{parallel extract} operation~$\pext(a,b)$ takes a bit mask~$a$ indicating a possibly non-consecutive bit field and extracts those bits from~$b$, packing them into ~$\popcount(a)$ bits.
  This corresponds to the~\texttt{pext} instruction on recent x86~processors.
  The operation is perhaps best understood with a diagram:
\begin{equation}
\begin{tabular}{rl}
$a$         &\texttt{1010111011000100}\\
$b$         &\texttt{1000101011110001}\\
bit field   &\texttt{1-0-101-11-{}-{}-0-{}-}\\
$\pext(a,b)$&\texttt{00000000\underline{10101110}}\\
\end{tabular}
\end{equation}
\item[pdep] The \emph{parallel deposit} operation~$\pdep(a,b)$ takes a bit mask~$a$ indicating a possibly non-consecutive bit field and deposits the bits from~$b$ into this field.
  It performs the opposite operation to~$\pext$ and corresponds to the~\texttt{pdep} instruction on recent x86~processors.
  We can likewise visualize its operation through a diagram:
\begin{equation}
\begin{tabular}{rl}
$a$         &\texttt{1010111011000100}\\
$b$&\texttt{10110100\underline{10101110}}\\
bit field   &\texttt{1-0-101-11-{}-{}-0-{}-}\\
$\pdep(a,b)$&\texttt{1000101011000000}\\
\end{tabular}
\end{equation}
\item[compress] The \emph{compress vector} operation~$\compress(m, v)$ is the only vector operation among our special functions.
  It performs the same operation as the parallel extract operation~$\pext$, 
  but instead of extracting bits from a bit field, it extracts elements from a vector.
  This corresponds to the \texttt{vpcompressb} instruction on recent x86 processors.
  For the visualization, we have given the mask~$m=\texttt{0xcd}$ with the least significant bit on the left to make the operation easier to see.
  The least significant mask bit decides whether to keep the first vector element and so on until the most significant mask bit decides whether to keep the last vector element:
\begin{equation}
\def\0{\hphantom0}
\begin{tabular}{rl}
$m$            &\texttt{\01\,\00\,\01\,\01\,\00\,\00\,\01\,\01}\\
$v$            &\texttt{12\,34\,56\,78\,9a\,bc\,de\,f0}\\
kept elements  &\texttt{12\,-{}-\,56\,78\,-{}-\,-{}-\,de\,f0}\\
$\compress(m,v)$&\texttt{\underline{12\,56\,78\,de\,f0}\,00\,00\,00}\\
\end{tabular}
\end{equation}
Observe how we reversed the bit order of the mask~$m$ to match the natural vector order: its usual binary representation is~\texttt{11001101}.
\end{description}

\begin{table}[tbh!]
\centering
\caption{Selected AVX-512 instructions.}  \label{table:ref:simdinstructions}
\setlength\tabcolsep{.5em}
\begin{tabular}{rcp{2.9in}}
\toprule
\emph{instruction} & {}\hskip-1em\emph{extension}\hskip-1em{} & \emph{description} \\
\midrule
\texttt{vmovdqu8/16}&BW&move byte/word/dword vector\\
\texttt{vpblendmw/d}&BW\!/F&blend words/dwords with mask\\
\texttt{vpbroadcastd/q}&F&broadcast dword/qword to vector\\
\texttt{vextracti32x8}&DQ&extract 256-byte word from vector\\
\texttt{vpmovzxbw/wd}&BW\!/F&zero-extend byte to word or word to dword\\
\texttt{vpaddb/w/d}&BW&add bytes/words/dwords\\
\texttt{vpsubb/w/d}&BW&subtract bytes/words/dwords\\
\texttt{vpcmpub/w}&BW&compare unsigned bytes/words\\
\texttt{vpternlogd}&F&logic on 3 operands by given truth table\\
\texttt{vpandd}&F&bitwise and dwords\\
\texttt{vpandnd}&F&bitwise and-not dwords\\
\texttt{vpsllw/d}&BW\!/F&logically shift words/dwords left by immediate\\
\texttt{vpsrlw/d}&BW\!/F&logically shift words/dwords right by immediate\\
\texttt{valignd}&F&right-shift elements between operands\\
\texttt{vpmultishiftqb}&VBMI&shift bytes within qword, see \S~\ref{16to8:encode}\\
\texttt{vpcompressb}&VBMI2&compress byte vector, see \S~\ref{notation:special} \\
\texttt{vpermb}&VBMI&permute byte vector by byte index vector\\[0.5em]
\texttt{kmovd/q}&BW&move 32/64-bit mask\\
\texttt{kord/q}&BW&bitwise or 32/64-bit mask\\
\texttt{kandnd/q}&BW&bitwise and-not 32/64-bit mask\\
\texttt{knotd/q}&BW&bitwise complement 32/64-bit mask\\
\texttt{kshiftrd/q}&BW&logically shift 32/64-bit mask right by imm.\\
\texttt{ktestd/q}&BW&test bitwise and/and-not of masks for all-zero\\
\texttt{kortestd/q}&BW&test bitwise or of masks for all-zero/all-one\\
\bottomrule
\end{tabular}
\end{table}

\section{AVX-512}
Our algorithms are based on the AVX-512 family of instruction-set extensions to the Intel~64\footnote{The 64 bit variant of the x86 (IA-32) instruction-set architecture, also known as \uc{amd64}, x86-64, \uc{em64t}, and IA-32e.} instruction-set architecture~\cite{sdm2}.
An extension to the AVX~family of instruction-set extensions, AVX-512 provides a comprehensive set of SIMD~instructions for operation on vectors of 16, 32, or 64~bytes organized into bytes or words of 16, 32, or 64~bits.  A register file of 32~\emph{vector registers} \texttt{zmm0}--\texttt{zmm31} complemented by 8~\emph{mask registers} \texttt{k0}--\texttt{k7} is provided.

AVX-512 instructions are generally non-destructive, writing their output into a separate operand from their inputs.
In most AVX-512 instructions, one operand is permitted to be a memory operand with the remaining operands being register or immediate operands.
This is usually the first input operand, but for some instructions it may also be the output operand.

The AVX-512 instruction set is split into a set of extensions.
Each extension adds new instructions to the Intel~64 architecture, enhancing the capabilities of AVX-512.
Depending on the microarchitecture used, not all AVX-512 extensions might be available.
Table~\ref{table:ref:simdinstructions} gives a list of AVX-512 instructions used and the extension they hail from.
In the following, we list those AVX-512 extensions needed to execute the algorithms described in this paper:
\begin{description}
\item[AVX-512F] The \emph{foundation} extension implements the basic AVX-512 instruction set on 64-byte vectors.
 Every AVX-512 implementation must support AVX-512F\@.
\item[AVX-512BW] The \emph{byte/word} extension extends the AVX-512F instructions to vectors of bytes and 16-bit words.
\item[AVX-512DQ] The \emph{dword/qword} extension provides additional instructions on 32- and 64-bit words.
\item[AVX-512VBMI] The \emph{vector byte manipulation instructions} extension adds instructions to permute and manipulate bytes.
\item[AVX-512VBMI2] The \emph{vector byte manipulation instructions~2} extension adds compress/expand support and double-width shifts for bytes and 16-bit words.
\end{description}
The first generation of Intel~64 processors supporting all required AVX-512 extensions are those code named~\emph{Icelake}, based on the microarchitecture code named~\emph{Sunny Cove}.
By emulating \texttt{vpcompressb} through other instructions, it is likely possible to adapt the algorithms to processors as early as the generation code named~\emph{Cannon Lake}, albeit at significant reduction in performance.

\subsection{Masking}
The output of most vector instructions is subject to \emph{masking}, a novel feature of AVX-512.
A mask register~\texttt{k1}--\texttt{k7}\footnote{Mask register~\texttt{k0} cannot be used for masking, but remains available for logic on masks.} is applied to the output operand, specifying either \emph{merge masking} or \emph{zero masking}.
With merge masking, only those vector elements indicated by bits set in the mask register are modified in the output operand.
The other vector elements remain unchanged.
With zero masking, vector elements for which the bits in the mask register are clear are zeroed out.

For example, the merge and zero masking instructions
\begin{equation*}
\begin{tabular}{ll}
\texttt{vpaddb zmm0\string{k1\string}, zmm2, zmm3}&(merge masking), and\\
\texttt{vpaddb zmm4\string{k5\string}\string{z\string}, zmm6, zmm7}&(zero masking)\\
\end{tabular}
\end{equation*}
perform a \textbf packed \textbf{add}ition of \textbf bytes, giving
\begin{align*}
\texttt{zmm0}&=\texttt{k1}\mathrel?\texttt{zmm2}+\texttt{zmm3}:\texttt{zmm0}\quad\mbox{and}\\
\texttt{zmm4}&=\texttt{k5}\mathrel?\texttt{zmm6}+\texttt{zmm7}:0.
\end{align*}
Masking on register operands is free for most instructions, though merge masking introduces an input dependency on the old value of the output operand.

Masking on memory operands enables \emph{memory fault suppression} for most instructions.
This means that the CPU does not signal memory faults for masked-out vector elements, permitting masked out elements to extend into unmapped or non-writable pages.
This suppression affects both input and output memory operands.

\subsection{Microarchitectural Details} \label{avx512:perf}
To simplify the implementation of AVX-512 on microarchitectures designed to execute the older SSE and AVX families of instruction-set extensions, most SIMD instructions operate within \emph{lanes} of 16~bytes.
That is, in many ways, it is as if the 64-byte vector registers were made of four nearly independent 16-byte subregisters.
Instructions that process data across lanes (such as~\texttt{vpermb} or~\texttt{vpcompressb}) exist, but can typically execute on less execution units and take longer to execute in comparison to instructions that do not.
We thus want to avoid cross-lane operations if feasible.

On current Intel microarchitectures including Sunny Cove (Icelake), Cypress Cove (Rocket Lake), and Willow Cove (Tiger Lake), most AVX-512 instructions\footnote{assuming no memory operands} can execute on \emph{execution ports}~0, 1, and~5.
Instructions that do not cross lanes usually execute in a single cycle, instructions that do take 3 or more cycles.
Some instructions are restricted in the ports they can execute on: shifts can only execute on ports~0/1, permutations and other cross-lane instructions, as well as comparisons into masks can only execute on port~5.
Instructions operating on masks (i.\,e.~those whose mnemonics start with~\texttt k) are restricted to one of ports~0 or port~5, depending on the instruction~\cite{agner4,Abel19a}.

In addition to these restrictions, ports 0 and~1 support a vector length of only 32~bytes while port~5 supports the whole 64~bytes.
Instructions operating on a vector length of 64~bytes are executed either on port~5 or on ports~0/1 joined together, occupying both ports for one cycle simultaneously.
Thus, there are effectively only two ports available to execute instructions with a 64-byte vector length.
While 32-byte vectors are processed at 3~vectors of 32~bytes (i.\,e. 6~lanes) per cycle, 64-byte vectors are processed at only 2~vectors of 64~bytes (or 8~lanes) per cycle, leading to a theoretical speedup by a factor of~$4/3$ or~$33\,\%$ of 64-byte vectors over 32-byte vectors for an otherwise identical algorithm.
This stands in contrast to the factor~$2$ or~$100\,\%$ speedup one would na\"ively expect from doubling the vector length.

It is vital for the performance of AVX-512 code to keep track of which ports instructions execute on, rearranging or editing the code such that both port~0/1 and port~5 can execute instructions at the same time~\cite{agner3}.
Through the use of microarchitectural simulation~\cite{Abel22} in the design of the algorithms, good port utilization has been ensured.

\section{Transcoding from UTF-8 to UTF-16} \label{sec:8to16}
We transcode UTF-8 to UTF-16 by gathering the bytes that make up each character from the last byte of each character to its first byte.
This exploits the similarity in bit arrangement between the four cases (ASCII, 2-byte, 3-byte, and 4-byte) highlighted in Fig.~\ref{16to8:correspondence}.
The bytes of each UTF-8 sequence are isolated from the input string, liberated of their tag bits, shifted into position, and finally summed up into a code point.

Using the exact correspondence between 4-byte UTF-8 characters and characters represented as surrogate pairs in UTF-16, we treat 4-byte characters as an overlapping pair of a 3-byte sequences and a 2-byte sequence that is later fixed up into a high and a low surrogate.
This saves us extra code for extracting the fourth-last byte of each sequence and avoids the costly use of 32-bit words for intermediate results.

To illustrate this idea, consider the following example, translating the Unicode characters \codepoint{40}~(@), \codepoint{A7}~(\S), \codepoint{2208}~($\in$), and \codepoint{1D4AA}~($\mathcal O$) from UTF-8 to UTF-16:
\begin{equation}
\setlength\tabcolsep{2pt}
\begin{tabular}{cccccccccc}
@&\multicolumn{2}{c}{\S}&\multicolumn{3}{c}{$\in$}&\multicolumn{4}{c}{$\mathcal O$}\\
\texttt{40}&\texttt{C2}&\texttt{A7}&\texttt{E2}&\texttt{88}&\texttt{88}&\texttt{F0}&\texttt{9D}&\texttt{92}&\texttt{AA}\\
\midrule
\texttt{40}&&&&&&&&&\\
&\texttt{C2}&\texttt{A7}&&&&&&&\\
&&&\texttt{E2}&\texttt{88}&\texttt{88}&&&&\\
&&&&&&\texttt{F0}&\texttt{9D}&\texttt{92}&\\
&&&&&&&&\texttt{92}&\texttt{AA}\\
\midrule
\multicolumn{2}{c}{\texttt{0040}}&\multicolumn{2}{c}{\texttt{00A7}}&\multicolumn{2}{c}{\texttt{2208}}&\multicolumn{2}{c}{\texttt{D835}}&\multicolumn{2}{c}{\texttt{DCAA}}\\
\end{tabular}
\end{equation}
These four characters demonstrate the behavior of the algorithm on the four UTF-8 cases, representing ASCII, 2-byte, 3-byte, and 4-byte respectively.
Observe especially how the code sequence~\texttt{F0 9D 92 AA} for~$\mathcal O$ is split into two overlapping sequences~\texttt{F0 9D 92} and~\texttt{92 AA}.
The first of these two is translated into the high surrogate~\texttt{D835} with the second one becoming the low surrogate~\texttt{DCAA}.

The algorithm can be roughly described with the following \emph{plan of attack}:

\begin{enumerate}
\item Read a vector of 64~bytes.
\item Classify each byte according to whether it is an ASCII byte, continuation byte, 2-byte lead byte, 3-byte lead byte, or 4-byte lead byte.
\item Construct a mask indicating the last byte of each UTF-8 sequence.  For 4-byte characters, the third byte is indicated, too, treating them as a 3-byte sequence for the high surrogate and a 2-byte sequence for the low surrogate.
\item Use the mask to gather the last, $2^{\mathrm{nd}}$~last, and $3^{\mathrm{rd}}$~last byte of each sequence.
\item Strip tag bits, shift bits into place and or them into UTF-16 words.
\item Postprocess surrogates by shifting their bits into place, and applying tag bits and surrogate plane shift.
\item Write the resulting bytes to the output, incrementing the input and output pointers by the number of bytes consumed/generated.
\item Repeat until the end of input or an encoding error are encountered.
\end{enumerate}

Apart from this general plan, there are also fast paths for the cases (a)~ASCII characters only, (b)~ASCII,  and 2-byte sequences only, and (c)~1--3-byte sequences only.

Validation is performed throughout the transcoding process, as explained in~\S~\ref{8to16:validation}.
In comparison to previous algorithms, it is simplified by advancing the input only by complete UTF-8 sequences;
if the input is correct UTF-8, each vector of input thus begins with a complete sequence.

\subsection{Classification and Masks}
After reading a vector of bytes from the input buffer, the characters in it are classified according to the range they fall into.  Various masks are then built from this classification.  In the following explanations, we follow the convention from~\S~\ref{notation} where names of the form~$m_{\ldots}$ refer to masks about the input vector~$w_{\mathrm{in}}$ while names of the form $M_{\ldots}$ refer to masks about the output vector.

These two kinds of masks are connected through the~$\pext$ and~$\pdep$ operations, relating the end bytes of the decoded UTF-8 sequences to the UTF-16 words they correspond to and vice versa.

The first set of masks is derived directly from~$w_{\mathrm{in}}$, classifying the input into ASCII
\begin{equation}
m_1=(w_{\mathrm{in}}<\texttt{0x80}), \label{8to16:m1}
\end{equation}
2/3/4-byte sequence lead bytes
\begin{equation}
m_{234}=(\texttt{0xc0}\le w_{\mathrm{in}}), \label{8to16:m234}
\end{equation}
3/4-byte sequence lead bytes
\begin{equation}
m_{34}=(\texttt{0xe0}\le w_{\mathrm{in}}) \label{8to16:m34}
\end{equation}
and 4-byte sequence lead bytes
\begin{equation}
m_4=(\texttt{0xf0}\le w_{\mathrm{in}}). \label{8to16:m4}
\end{equation}
From these we then derive a mask
\begin{equation}
m_{1234}=m_1\lor m_{234} \label{8to16:m1234}
\end{equation}
indicating the presence of any kind of lead byte.
All other bytes ($\lnot m_{1234}$) are continuation bytes.

Then we construct the important mask~$m_{\mathrm{end}}$ identifying the last bytes of each sequence to be decoded.
These are the last bytes of each UTF-8 sequence as well as the third byte of each 4-byte sequence.
Working backwards from these last bytes, we later use this mask to gather the last, second-last and third-last bytes of each sequence.

The key insight in constructing $m_{\mathrm{end}}$ is that as each UTF-8 sequence is followed by another UTF-8 sequence, we can find the positions of the last bytes as those preceding the lead bytes of the next sequence~($m_{1234}\gg1$).
The third byte of each 4-byte sequence is added by first computing the fourth byte of each sequence
\begin{equation}
m_{+3}=m_4\ll3 \label{8to16:mplus3}
\end{equation}
and then shifting the result to the right to obtain the last third bytes
\begin{align}
m_{\mathrm{end}}=(m_{+3}\lor m_{1234})\gg1\lor m_{+3}.\label{8to16:mend}
\end{align}

An unfortunate consequence of defining~$m_{\mathrm{end}}$ by going backwards from the lead bytes of the next characters is that we only catch the last character of the vector when it is followed by an incomplete character whose lead byte we can shift to the right. 
For 4-byte sequences right at the end of the vector, this leads to us only detecting the third last byte in the character.
Oring in~$m_{+3}$ at the end fixes this problem for the 4-byte case.

For the other cases, the only effect of this process is that if $w_{\mathrm{in}}$~does not end in a partial character, decodes to no more than 32~words of UTF-16, and the last character is not a 4-byte sequence, we process one less character in the current iteration than possible.
However, the minor performance impact of hitting this edge case is more than outweighed by not spending extra time computing the mask correctly.\footnote{
If a perfect mask is desired, you can instead use
\begin{equation*}
m'_{\mathrm{end}}=(m_1\lor m_2\ll 1\lor m_{34}\ll2\lor m_4\ll3)\land\lnot\bigl(m_4\ll2\land 1\ll(n-1)\bigr), \label{8to16:mendperfect}
\end{equation*}
where $m_2=m_{234}\land\lnot m_{34}$ indicates 2-byte sequence lead bytes.  
The first byte of each sequence is shifted to the position of its last byte (and the third byte of a 4-byte sequence).
The mask is then post-processed by clearing the third byte of a 4-byte sequence starting in the third-last byte of~$w_{\mathrm{in}}$, as only complete sequences can be processed.}

To visualize the various masks, consider the strings ``x$\nabla\mathfrak P$'' and ``$\varepsilon{\le}{\pm}1$'' with a vector length of~$n=8$~bytes:
\begin{equation}
\setlength\tabcolsep{2pt}
\begin{tabular}{rlccccccccclcccccccc}
&&&x&\multicolumn{3}{c}{$\nabla$}&\multicolumn{4}{c}{$\mathfrak P$}&&\multicolumn{2}{c}{$\varepsilon$}&\multicolumn{3}{c}{$\le$}&\multicolumn{2}{c}{$\pm$}&$1$\\
$w_{\mathrm{in}}$&&\kern.5em&\texttt{78}&\texttt{e2}&\texttt{88}&\texttt{87}&\texttt{f0}&\texttt{9d}&\texttt{94}&\texttt{93}&\kern1em&\texttt{ce}&\texttt{b5}&\texttt{e2}&\texttt{89}&\texttt{a4}&\texttt{c2}&\texttt{b1}&\texttt{31}\\
\midrule
$m_1$     &$\!=w_{\mathrm{in}}<\texttt{0x80}$   &&1&0&0&0&0&0&0&0&&0&0&0&0&0&0&0&1\\
$m_{234}$ &$\!=\texttt{0xc0}\le w_{\mathrm{in}}$&&0&1&0&0&1&0&0&0&&1&0&1&0&0&1&0&0\\
$m_{34}$  &$\!=\texttt{0xe0}\le w_{\mathrm{in}}$&&0&1&0&0&1&0&0&0&&0&0&1&0&0&0&0&0\\
$m_4$     &$\!=\texttt{0xf0}\le w_{\mathrm{in}}$&&0&0&0&0&1&0&0&0&&0&0&0&0&0&0&0&0\\
$m_{1234}$&$\!=m_1\lor m_{234}$                 &&1&1&0&0&1&0&0&0&&1&0&1&0&0&1&0&1\\
$m_{+3}$  &$\!=m_4\ll3$                         &&0&0&0&0&0&0&0&1&&0&0&0&0&0&0&0&0\\
$m_{\mathrm{end}}$&$\!=(m_{+3}\lor m_{1234})\gg1\lor m_{+3}$&&1&0&0&1&0&0&1&1&&0&1&0&0&1&0&1&0\\
\end{tabular}
\end{equation}
In this example, masks are arrays of eight bits corresponding to eight-byte sequences: in our actual implementation, we use 64-bit masks.
Note in particular how $m_{\mathrm{end}}$ accounts for the last character in the left string (being a 4-byte character), but not in the right string, where it is an ASCII character.
Also note how the character~$\mathfrak P$ has two end bits, being treated as a 3-byte sequence overlapping a 2-byte sequence.

\subsection{Assembling Characters}
\label{sec:assembling}
With these masks in hand, we can strip off the tag bits and assemble characters.
The UTF-8 tag bits are stripped off by clearing the most significant two bit of each non-ASCII byte in~$w_{\mathrm{in}}$, giving
\begin{equation}
w_{\mathrm{stripped}}=m_1\mathrel ? w_{\mathrm{in}}:w_{\mathrm{in}}\land\texttt{0x3f}. \label{8to16:wstripped}
\end{equation}
The tag bits of 3/4-byte lead bytes are not completely removed by this step;
this is sufficient for our purposes as these tag bits  get shifted out later on.

Characters are assembled by selecting from~$w_{\mathrm{stripped}}$ the last~($W_{\mathrm{end}}$), second-last~($W_{-1}$) and third-last bytes~($W_{-2}$) of each sequence, zero-extending them to 16~bits and joining their bits into a UTF-16 word.
We do this by first preparing a permutation vector~$P$ that holds for each word in the output vector, the index of the last byte of the corresponding sequence.
This vector is prepared by compressing (\texttt{vpcompressb}) a byte vector holding an identity permutation~$(0,1,\ldots,63)$ subject to~$m_{\mathrm{end}}$.  The compressed vector is then zero-extended~(\texttt{vpmovzxbw}) to 16-bit words, keeping its first~$n/2$ elements:
\begin{equation}
P=\compress\bigl(m_{\mathrm{end}}, (0,1,\ldots,n-1)\bigr). \label{8to16:P}
\end{equation}
We only generate one vector of UTF-16 words per iteration representing at most 32~characters. When the input contains ASCII characters, it might be possible for $m_{\mathrm{end}}$ to contain more than 32~set bits. Bits set in~$m_{\mathrm{end}}$ past the 32$^\mathrm{nd}$~bit are discarded during the processing: $P$ contains only $n/2$ (or 32) elements.
With $P$ in hand, we can load the last byte of each sequence
\begin{equation}
W_{\mathrm{end}}[i]=w_{\mathrm{stripped}}\bigl[P[i]\bigr] \label{8to16:Wend}
\end{equation}
with a single permutation instruction (\texttt{vpermb}).\footnote{as \texttt{vpermb} permutes each byte, we zero-mask its result with \texttt{0x5555555555555555} to only permute into the less significant byte of each 16-bit word, zero-extending for free.}

By decrementing the entries of~$P$, we produce index vectors corresponding to the second-last and third-last bytes of each sequence.
To avoid loading the third-last byte of a 1/2-byte sequence or the second-last byte of an ASCII sequence, we mask~$w_{\mathrm{stripped}}$ with masks
\begin{align}
m_{-1}&=\lnot m_1\gg 1\quad\mbox{and} \label{8to16:mminus1} \\
m_{-2}&=m_{34}\land\lnot0\gg2 \label{8to16:mminus2}
\end{align}
to clear out bytes before ASCII characters resp.\ those that do not start a 3/4-byte sequence, accounting for possible wrap around.\footnote{$P[i]-1$ and~$P[i]-2$ may yield negative numbers; we assume that in a permutation, such indices either wrap around to the end of the vector or produce~$0$ as an output. If negative permutation indices yield zeroes, the term~$\lnot0\gg2$, serving as wraparound protection, can be omitted from~$m_{-2}$.}
We then obtain our vectors
\begin{align}
W_{-1}[i]&=\bigl(m_{-1}\mathrel?w_{\mathrm{stripped}}:0\bigr)\bigl[P[i]-1\bigr]\quad\mbox{and} \label{8to16:Wminus1}\\
W_{-2}[i]&=\bigl(m_{-2}\mathrel?w_{\mathrm{stripped}}:0\bigr)\bigl[P[i]-2\bigr] \label{8to16:Wminus2}
\end{align}
as desired.
The last, second-last, and third-last bytes are shifted into place and ored such that the bits~\texttt A--\texttt W
are contiguous, giving
\begin{equation}
W_{\mathrm{sum}}=W_{-2}\ll12\lor W_{-1}\ll6\lor W_{\mathrm{end}}.  \label{8to16:Wsum}
\end{equation}
The value of~$W_{\mathrm{sum}}$ depending on the case taken can be visualized as follows:
\begin{equation}
\begin{tabular}{rcc}
\emph{case}&\emph{byte sequence}&$W_{\mathrm{sum}}$\\
\midrule
ASCII   &\texttt{\hphantom{XXXX\,XXXX\;XXXX\,XXXX\;XXXX\,XXXX\;}0GFE\,DCBA}&\texttt{0000\,0000\,0GFE\,DCBA}\\
2 byte  &\texttt{\hphantom{XXXX\,XXXX\;XXXX\,XXXX\;}110L\,KJHG\;10FE\,DCBA}&\texttt{0000\,0LKJ\,HGFE\,DCBA}\\
3 byte  &\texttt{\hphantom{XXXX\,XXXX\;}1110\,RQPN\;10ML\,KJHG\;10FE\,DCBA}&\texttt{RQPN\,MLKJ\,HGFE\,DCBA}\\
hi surr &\texttt{1111\,0WVU\;10TS\,RQPN\;10ML\,KJHG\hphantom{\;XXXX\,XXXX}}&\texttt{0WVU\,TSRQ\,PNML\,KJHG}\\
lo surr &\texttt{\hphantom{XXXX\,XXXX\;XXXX\,XXXX\;}10ML\,KJHG\;10FE\,DCBA}&\texttt{0000\,MLKJ\,HGFE\,DCBA}\\
\end{tabular}
\end{equation}

This representation is close to UTF-16LE format with only the surrogate cases diverging.
To address this difference, we first identify the locations of surrogates in~$W_{\mathrm{out}}$.
Sequences in~$w_{\mathrm{in}}$ corresponding to low surrogates end at the fourth bytes of 4-byte sequences.
By extracting the locations of these through~$m_{\mathrm{end}}$ into the space of~$W_{\mathrm{out}}$, we obtain the locations of low surrogates
\begin{equation}
M_{\mathrm{lo}}=\pext(m_{\mathrm{end}}, m_{+3}) \label{8to16:Mlo}
\end{equation}
in~$W_{\mathrm{out}}$.
High surrogates
\begin{equation}
M_{\mathrm{hi}}=M_{\mathrm{lo}}\gg1 \label{8to16:Mhi}
\end{equation}
always precede low surrogates.

Surrogates are fixed up by shifting high surrogates into position and applying surrogate plane shift and tag bits,\footnote{Adding $\texttt{0xd7c0}=\texttt{0xd800}-\texttt{0x0020}$ applies the tag bits and the surrogate plane shift in one step.} giving
\begin{equation}
W_{\mathrm{out}}=
\begin{cases}
(W_{\mathrm{sum}}\gg4)+\texttt{0xd7c0}&\mbox{if $M_{\mathrm{hi}}$}\\
W_{\mathrm{sum}}\lor \texttt{0xdc00}&\mbox{if $M_{\mathrm{lo}}$}\\
W_{\mathrm{sum}}&\mbox{otherwise.}
\end{cases} \label{8to16:Wout}
\end{equation}
The operation of Eq.~\ref{8to16:Wout} can be visualized as follows, where $\texttt{0vuts}=\texttt{WVUTS}-1$:
\begin{equation}
\begin{tabular}{rcc}
\emph{case}&$W_{\mathrm{sum}}$&$W_{\mathrm{out}}$\\
\midrule
high surrogate&\texttt{0WVU\,TSRQ\,PNML\,KJHG}&\texttt{1101\,10vu\,tsRQ\,PNML}\\
low surrogate&\texttt{0000\,MLKJ\,HGFE\,DCBA}&\texttt{1101\,11KJ\,HGFE\,DCBA}\\
other&\texttt{RQPN\,MLKJ\,HGFE\,DCBA}&\texttt{RQPN\,MLKJ\,HGFE\,DCBA}\\
\end{tabular}
\end{equation}
For illustration purposes, we provide C~code implementing
equation Eq.~\ref{8to16:Wout} using Intel intrinsic functions: see Fig.~\ref{lst:tab-new-line-cpp}.

\begin{figure}
\centering
\begin{tabular}{c} 
\begin{lstlisting}[style=customcpp]
__m512i mask_d7c0d7c0 = _mm512_set1_epi32(0xd7c0d7c0);
__m512i mask_dc00dc00 = _mm512_set1_epi32(0xdc00dc00);
//...
// Mlo, Mhi and Wsum have been computed, we compute Wout.
__m512i lo_surr_mask = _mm512_maskz_mov_epi16(Mlo, mask_dc00dc00);
__m512i shifted4_Wsum = _mm512_srli_epi16(Wsum, 4);
__m512i tagged_lo_surrogates = _mm512_or_si512(Wsum, lo_surr_mask); 
__m512i Wout = _mm512_mask_add_epi16(tagged_lo_surrogates, Mhi, 
               shifted4_Wsum, mask_d7c0d7c0);
\end{lstlisting}
  \end{tabular}
  \caption{C code using Intel intrinsic functions equivalent to Eq.~\ref{8to16:Wout}.}\label{lst:tab-new-line-cpp}
\end{figure}

The vector~$W_{\mathrm{out}}$ holds the UTF-16LE encoded characters we want to write out.  
There is a final issue: the 64~bytes of UTF-8 data in the input may correspond to anywhere from 21 to~64 words of output, of which the first up to 32~words are processed.\footnote{Input data corresponding to the remaining words (if any) is reprocessed in the next iteration.}
If a surrogate pair happened to straddle the end of~$W_{\mathrm{out}}$, we would discard the corresponding low surrogate and produce an incorrect result.
So once again, special care must be taken to omit the $32^{\mathrm{nd}}$~word of output if it is a high surrogate.
We do so by computing a mask
\begin{equation}
M_{\mathrm{out}}=\lnot\bigl(M_{\mathrm{hi}}\land 1\ll(n/2-1)\bigr) \label{8to16:Mout}
\end{equation}
of the elements of~$W_{\mathrm{out}}$ excluding the last element if it happens to be a high surrogate.
We introduce a variable~$b$ which is set to all ones  ($b=\lnot0$) except at the end of the input (cf.~\S~\ref{8to16:tail}).
By depositing the mask $M_{\mathrm{out}}$ into the last bytes of each sequence, we obtain a mask
\begin{equation}
m_{\mathrm{processed}}=\pdep(b\land m_{\mathrm{end}}, M_{\mathrm{out}}) \label{8to16:mprocessed}
\end{equation}
holding the locations of the last byte of each sequence that has been processed into a word in~$W_{\mathrm{out}}$.

With this mask, we can compute the number of bytes of input processed
\begin{equation}
n_{\mathrm{in}}=\width(m_{\mathrm{processed}}) \label{8to16:nin}
\end{equation}
and the number of words of output produced
\begin{equation}
n_{\mathrm{out}}=\popcount(m_{\mathrm{processed}}). \label{8to16:nout}
\end{equation}
The first $n_{\mathrm{out}}$~bytes of the output vector are then deposited into the output buffer, input and output buffers are advanced by~$n_{\mathrm{in}}$ and~$n_{\mathrm{out}}$ and we continue with the next iteration.

To visualize the generation of~$m_{\mathrm{processed}}$, consider the example string ``$\pm1{=}\mathcal O$'' with a vector length of~$n=8$~bytes:
\begin{equation}
\setlength\tabcolsep{2pt}
\begin{tabular}{rcccccccc}
&\multicolumn{2}{c}{$\pm$}&1&=&\multicolumn{4}{c}{$\mathcal O$}\\
$w_{\mathrm{in}}$&\texttt{C2}&\texttt{B1}&\texttt{31}&\texttt{3D}&\texttt{F0}&\texttt{9D}&\texttt{92}&\texttt{AA}\\
$m_{\mathrm{end}}$&0&1&1&1&0&0&1&1\\
\end{tabular}
\end{equation}
As the character \texttt{D835 DCAA} straddles the end of the vector, it cannot be processed in the current iteration:
\begin{equation}
\setlength\tabcolsep{2pt}
\begin{tabular}{rccccc}
&$\pm$&1&=&\multicolumn{2}{c}{$\mathcal O$\hphantom{()}}\\
$W_{\mathrm{out}}$&\texttt{00B1}&\texttt{0031}&\texttt{003D}&\texttt{D835}&(\texttt{DCAA})\\
$M_{\mathrm{hi}}$&0&0&0&1\\
$1\ll(n/2-1)$&0&0&0&1\\
$M_{\mathrm{out}}$&1&1&1&0
\end{tabular}
\end{equation}
Depositing the bits of~$M_{\mathrm{out}}$ through~$m_{\mathrm{end}}$, we then obtain
\begin{equation}
\setlength\tabcolsep{2pt}
\begin{tabular}{rcccccccc}
$m_{\mathrm{processed}}$&0&1&1&1&0&0&0&0\\
bytes processed&\texttt{C2}&\texttt{B1}&\texttt{31}&\texttt{3D}&\texttt{-{}-}&\texttt{-{}-}&\texttt{-{}-}&\texttt{-{}-}\\
words produced&\multicolumn{2}{c}{\texttt{00B1}}&\multicolumn{2}{c}{\texttt{0031}}&\multicolumn{2}{c}{\texttt{003D}}&\multicolumn{2}{c}{\texttt\dash}\\
\end{tabular}
\end{equation}
and advance buffers by $n_{\mathrm{in}}=4$~bytes and $n_{\mathrm{out}}=3$~words respectively.
The bytes corresponding to~$\mathcal O$ will be processed again in the next iteration.

\subsection{Processing the Tail} \label{8to16:tail}
The final bit of input with less than 64~characters remaining (tail) is handled through the variable~$b$.
This variable holds a mask of those bytes in~$w_{\mathrm{in}}$ we are permitted to process.
Initially we set~$b=\lnot0$, permitting all bytes to be processed.
When the end of the input with~$\ell<n$ bytes remaining to be processed is reached, we set $b$~to a mask of the first $\ell$~bytes of~$w_{\mathrm{in}}$, giving
\begin{equation}
b=(1\ll\ell)-1. \label{8to16:b}
\end{equation}
The tail of input is read zero-masked by~$b$, padding it with \uc{nul}~bytes.
Then, a final iteration of the main loop is performed, processing only the bytes accounted for in~$b$.

\subsection{Input Validation} \label{8to16:validation}
Throughout the transcoding process, we check the input for encoding errors and abort transcoding if any such error occurs.
Aborting is done by determining the location of the encoding error and setting the remaining input length~$\ell$ to the number of bytes preceding the first error.
We then clear all input bytes starting at the first erroneous byte and jump to the tail-handling code from~\S~\ref{8to16:tail}, effectively restarting the current iteration as ``final'' iteration.

Having talked about how to continue after an error has occurred, we shall now direct our attention to the kinds of errors we have to check for.
A UTF-8 encoded document must conform to the following rules:

\begin{enumerate}
\item Bytes \texttt{0xf5}--\texttt{0xff} must not occur.
\item Lead and continuation bytes must match: each byte in \texttt{0xc0} to~\texttt{0xdf} must be followed by one continuation byte, each byte from \texttt{0xe0} to~\texttt{0xef} by two continuation bytes and each byte from \texttt{0xf0} to~\texttt{0xf4} by three continuation bytes.
\item Continuation bytes may not otherwise occur.
\item The decoded character must be larger than~\codepoint{7F} for 2-byte sequences, larger than~\codepoint{7FF} for 3-byte sequences, and larger than~\codepoint{FFFF} for 4-byte sequences. 
\item The character must be no greater than \codepoint{10FFFF}. 
\item The character must not be in the range \codepointrange{D800}{DFFF}. 
\end{enumerate}

We check for these rules throughout the algorithm, mostly reusing masks we already have to compute for other steps of the code.
Three checks are performed in total:

\paragraph{Overlong 2-byte sequences}
Right at the beginning, we check whether any of the bytes \texttt{0xc0} or~\texttt{0xc1} occur.
Presence of these bytes indicates a 2-byte sequence that encodes a code point below~\codepoint{80}, violating condition~4.
The first invalid input byte is the first \texttt{0xc0}~or \texttt{0xc1}~byte found:
\begin{equation}
\mbox{valid if}\quad\bigl(m_{234}\land(w_{\mathrm{in}}<\texttt{0xc2})\bigr)=0.\label{8to16:fail1}
\end{equation}

\paragraph{Mismatched continuation bytes}
After computing the various classification masks, we check if conditions~2 and~3 hold.
As each byte of UTF-8 is either a lead or continuation byte, we check this by computing where continuation bytes should be~($m_c$) 
and comparing this with where lead bytes are not:
\begin{equation}
\mbox{valid if}\quad m_c=\lnot m_{1234}.\label{8to16:fail2}
\end{equation}
We compute~$m_c$ from the location of the second ($m_{+1}$), third~($m_{+2}$), and fourth byte ($m_{+3}$, see Eq.~\ref{8to16:mplus3}) of each sequence:
\begin{align}
m_{+1}&=m_{234}\ll1, \label{8to16:mplus1} \\
m_{+2}&=m_{34}\ll2, \label{8to16:mplus2} \\
m_c&=m_{+1}\lor m_{+2}\lor m_{+3}. \label{8to16:mc}
\end{align}

Conveniently, this check also fails on the input if it starts with continuation bytes, violating the invariant established earlier.
We do not catch a UTF-8 sequence straddling the end of the vector; such a sequence is checked properly in the next iteration once additional bytes have been fed in.

If this check fails, we must distinguish two cases to determine the location of the first encoding error:
If the first mismatch of $m_c$ and~$m_{1234}$ is due to a continuation byte present where there should not be one, the first invalid byte is that byte, giving
\begin{equation}
\ell=\ctz(m_c\oplus\lnot m_{1234}).\label{8to16:l1}
\end{equation}
Otherwise a continuation byte is missing where there should be one and the corresponding lead byte is the first invalid byte.
This byte can be found by masking~$m_{1234}$ to all bits preceding the mismatch
\begin{equation}
m_{\mathrm{pre}}=\bigl(1\ll\ctz(m_c\oplus\lnot m_{1234})\bigr)-1 \label{8to16:mpre}
\end{equation}
and then finding the last (most significant) bit in it, corresponding to the lead byte that is missing a continuation byte.  This gives
\begin{equation}
\ell=\width(m_{1234}\land m_{\mathrm{pre}})-1.\label{8to16:l2}
\end{equation}

\paragraph{Encodings out of range}
Finally, we check if the codepoints encoded by 3-~and 4-byte sequences are in range (conditions~4 and~5) and that 3-byte sequences do not encode surrogates (condition~6).
The algorithm treats input bytes in the range \texttt{0xf5}--\texttt{0xff} as lead bytes of 4-byte sequences.
Such sequences encode code points well in excess of~\codepoint{110000}, allowing us to verify condition~1 as a side effect with no extra code.

We augment our existing mask set with a mask
\begin{equation}
m_3=m_{34}\land\lnot m_4 \label{8to16:m3}
\end{equation}
indicating the location of 3-byte sequence start bytes in~$w_{\mathrm{in}}$.
Shifting the mask to indicate the last byte of each 3-byte sequence, extracting through~$m_{\mathrm{end}}$, and truncating to $n/2$~bits, we obtain a mask
\begin{equation}
M_3=\pext(m_{\mathrm{end}}, m_3\ll2) \label{8to16:M3}
\end{equation}
indicating which words in~$W_{\mathrm{out}}$ correspond to 3-byte sequences.
We then use~$M_3$ to check if any 3-byte sequences encode codepoints below~\codepoint{800},
\begin{equation}
M_{<\codepoint{800}}=M_3\land(W_{\mathrm{out}}<\texttt{0x800}) \label{8to16:M800}
\end{equation}
indicating violations of condition~4.

Then we check for surrogates: words in~$M_3$ must not encode surrogates, words in~$M_{\mathrm{hi}}$ must encode high surrogates (condition~6).\footnote{by construction, words produced from 1-~and 2-byte sequences never produce surrogates and $M_{\mathrm{lo}}$~always produces low surrogates; we do not need to validate these.}
A word in~$M_{\mathrm{hi}}$ produces a high surrogate if and only if the code point it encodes is in range \codepointrange{10000}{10ffff} (conditions~1, 4, and~5).
The masks
\begin{align}
M_{3s}&=M_3\land(\texttt{0xd800}\le W_{\mathrm{out}}<\texttt{0xe000}) \nonumber\\
&=M_3\land(W_{\mathrm{out}}-\texttt{0xd800}<\texttt{0x0800}) \label{8to16:M3s}
\end{align}
and
\begin{align}
M_{4s}&=M_{\mathrm{hi}}\land\lnot(\texttt{0xd800}\le W_{\mathrm{out}}<\texttt{0xdc00}) \nonumber\\
&=M_{\mathrm{hi}}\land(W_{\mathrm{out}}-\texttt{0xd800}\ge\texttt{0x0400}) \label{8to16:M4s}
\end{align}
indicate violations of these conditions.\footnote{As all comparisons are unsigned~(\texttt{vpcmpltuw}), one comparison for each range check suffices.}
The check succeeds if no offending words are found:
\begin{equation}
\mbox{valid if}\quad M_{<\codepoint{800}}\lor M_{3s}\lor M_{4s}=0.\label{8to16:fail3}
\end{equation}

If an offending word is found, the first invalid byte is the start byte of the corresponding sequence.
As the error can never occur in a low surrogate, we can find its location by projecting its location back onto the locations of the first and fourth bytes of every sequence:
\begin{equation}
\ell=\ctz\bigl(\pdep(m_{+3}\lor m_{1234},\;M_{<\codepoint{800}}\lor M_{3s}\lor M_{4s})\bigr).\label{8to16:l3}
\end{equation}

\subsection{Fast Paths}
Three fast paths are provided, speeding up common cases.  The first two are programmed such that they cannot be triggered in the ``final'' iterations for the tail or in case of an encoding error, allowing us to omit the handling of~$b$ in their length computations for a further performance increase.

\paragraph{ASCII only}
If the first 32~bytes of input are all ASCII bytes, we process these by  zero-extension (\texttt{vpmovzxbw}) of the first 32~bytes to 16-bit words.
The number of processed bytes is always~32, the number of words written out always~32, shortening the dependency chain to the next iteration.  No validation is needed in this case as ASCII bytes are always valid.

Only the first 32~bytes are considered before embarking on the fast path as the default path does not process more than 32~characters in any case.
Hence, while checking for all 64~bytes to be ASCII would allow for slightly faster processing in the all-ASCII case, performance for documents with short runs of ASCII characters amidst other characters (e.\,g.~HTML documents) suffers significantly, outweighing the benefits of the other case.

\paragraph{1/2 byte only}
In the absence of 3-~and 4-byte sequences ($m_{34}=0$), we employ a simplified variant of the algorithm.
While following the same operating principles as the main algorithm, we can take some shortcuts in the proven absence of 3-~and 4-byte sequences.  
First, the computation of some masks is greatly simplified, with most masks being entirely irrelevant for this path:
\begin{align}
m_2&=m_{234}, \label{8to16:12b:m2}\\
m_{\mathrm{end}}&=\lnot m_2,\quad\mbox{and} \label{8to16:12b:mend}\\
M_{\mathrm{out}}&=\pdep\bigl(m_{\mathrm{end}}, (1\ll n/2)-1\bigr). \label{8to16:12b:mout}
\end{align}

We then employ a simplified scheme to compute~$W_{\mathrm{out}}$:
Instead of masking out tag bits, we subtract~\texttt{0xc2} from the lead byte of each two-byte sequence to cancel out the tag bits of both lead and continuation byte, giving
\begin{equation}
w_{-\texttt{0xc2}}=m_1\mathrel?0:w_{\mathrm{in}}-\texttt{0xc2}. \label{8to16:wc2}
\end{equation}
Instead of first building a permutation vector~$P$ and then using it to permute the input bytes into place, we directly compress the bytes into position (\texttt{vpcompressb}) and then zero extend to 16-bit words (\texttt{vpmovzxbw}), giving
\begin{align}
W_{\mathrm{end}}&=\compress(m_{\mathrm{end}}, w_{\mathrm{in}})\quad\mbox{and} \label{8to16:12b:Wend}\\
W_{-1}&=\compress(m_{1234}, w_{-\texttt{0xc2}}). \label{8to16:12b:Wminus1}
\end{align}
Vectors $W_{\mathrm{end}}$ and~$W_{-1}$ must be merged by addition instead of bitwise or to correctly cancel out tag bits, giving
\begin{equation}
W_{\mathrm{out}}=(W_{-1}\ll6)+W_{\mathrm{end}}. \label{8to16:12b:Wout}
\end{equation}
The operation on 2-byte characters can be visualized as follows; \texttt{0xc2} is subtracted separately to illustrate the idea:
\begin{equation}
\begin{tabular}{rll}
&$W_{\mathrm{end}}$&\texttt{0000\,0000\;\underline{10FE\,DCBA}}\\
$+$&$W_{\mathrm{-1}}\ll6$&\texttt{00\underline{11\,0LKJ\;HG}00\,0000}\\
$-$&$\texttt{0xc2}\ll6$&\texttt{00\underline{11\,0000\;10}00\,0000}\\
\midrule
$=$&$W_{\mathrm{out}}$&\texttt{0000\,0LKJ\;HGFE\,DCBA}\\
\end{tabular}
\end{equation}
We want to increment the input pointer quickly---without a long chain of operations. We find it advantageous to always process half a vector (32~bytes or 33~bytes to include a final continuation byte) of input data per iteration like in the ASCII-only fast path.
While this approach usually processes less data than first determining the maximum number of input bytes we can process, being able to load the next data quicker is more important.  We avoid
accessing the SIMD masks to determine whether we advance
by 32~bytes or 33~bytes.


Thus we have
\begin{equation}
n_{\mathrm{in}}=\begin{cases}
n/2+1&\mbox{if $\texttt{0x80}\le w_{\mathrm{in}}[n/2]<\texttt{0xc0}$}\\
n/2&\mbox{otherwise,}
\end{cases}
\end{equation}
processing 32 bytes per iteration unless a 2-byte sequence straddles the middle of the vector\footnote{i.\,e. unless the byte at position~$n/2$ is a continuation byte}, in which case we process that extra byte, too.

The output buffer is advanced by the number of characters starting in the first 32~bytes, giving
\begin{equation}
n_{\mathrm{out}}=\popcount\bigl(m_{1234} \land (1\ll n/2)-1\bigr).
\end{equation}


As for validation, the checks for ``encodings out of range'' are omitted.  The check for ``mismatched continuation bytes'' is simplified to
\begin{equation}
\mbox{valid if}\quad m_2\ll1=\lnot m_{1234} \label{8to16:12b:fail2}
\end{equation}
as continuation bytes must always directly follow 2-byte sequence lead bytes.
The combination of all these simplifications yields a code path of roughly half the latency of the standard code path.

\paragraph{1/2/3 byte only}
In the absence of 4-byte sequences ($m_{4}=0$),
all characters are in the Basic Multilingual Plane. In this common case,
we can slightly simplify the main routine. We have
that $m_{+3}=m_4\ll3$ is zero. Consequently,
we can simplify the definitions of $m_c$ and $m_{\mathrm{end}}$ to
\begin{align}
m_c&=m_{+1}\lor m_{+2}\quad\mbox{and}\\
m_{\mathrm{end}}&=m_{1234}\gg1.
\end{align}
The computation of $W_{\mathrm{out}}$ and $M_\mathrm{out}$ is eliminated. As no surrogates are present, we can omit the surrogate post-processing and don't need to account for surrogate pairs straddling the end of the vector.
Instead, we directly get
\begin{align}
W_{\mathrm{out}}&=W_{\mathrm{sum}}\quad\mbox{and}\\
M_{\mathrm{out}}&=\lnot0.
\end{align}
Finally, the validation check for out-of-range encoding is slightly simpler: as surrogates cannot occur, we can drop the $M_{4s}$~term off Eq.~\ref{8to16:fail3}.

\section{Transcoding from UTF-16 to UTF-8}
As explained in~\S~\ref{unicode:utf-16}, UTF-16 encodes characters in the Basic Multilingual Plane~(\codepointrange{0000}{ffff}) in one 16-bit word and all others in two words as \emph{surrogate pairs}.
To encode a code~point as a surrogate pair, \texttt{0x10000} is subtracted from the character code to obtain a 20-bit binary number.
The most significant 10~bits are added to~\texttt{0xD800} to form a \emph{high surrogate}, which is followed by the less significant 10~bits added to~\texttt{0xDC00}, producing the corresponding \emph{low surrogate}.

UTF-8 encodes Unicode characters in the range \codepointrange{0000}{007F} in one byte, characters in the range \codepointrange{0080}{07FF} in two bytes, characters in the range \codepointrange{0800}{FFFF} in three bytes and the other characters in four bytes.
Characters encoded in one UTF-16 word thus correspond to characters encoded in 1--3~bytes of UTF-8 and characters encoded in two UTF-16 words correspond to characters encoded in 4~bytes of UTF-8.
This suggests the following \emph{plan of attack} for transcoding UTF-16 to UTF-8:

\begin{enumerate}
\item Read a vector of 16-bit words.
\item Classify the input words into ASCII (\hexrange{0000}{007F}), 2-byte (\hexrange{0080}{07FF}), high surrogate (\hexrange{D800}{DBFF}), low surrogate (\hexrange{DC00}{DFFF}), and 3-byte (\hexrange{0800}{FFFF}).
\item \emph{Zero extend} each 16-bit word to a 32-bit word and join low and high surrogates.
\item Shuffle the bits within each 32-bit word into the right positions and apply tag bits according to the type of character, producing UTF-8 sequences padded with null~bytes.
\item \emph{Compress} this vector, squeezing out the padding bytes.
\item Write the byte string to the output buffer and proceed to the next iteration.
\end{enumerate}

Apart from this general plan, we also have fast code paths for the three cases of (a)~ASCII characters only, (b)~all in \codepointrange{0000}{07FF}, and (c)~no surrogates, complementing the default code path (d)~surrogates present.  
Which code path to take is decided based on the characters in the current 62-byte chunk of input. 
We expect that  most text inputs would consistently rely on the same code paths. Thus branches corresponding to the various fast paths are easy to predict, and we expect that they may provide a significant performance boost.

We would now like to explain the steps in the \emph{plan of attack} in detail.  The steps are interlinked with information produced in each step being reused for the subsequent steps.  Additionally, the classification masks are reused for input validation.

First, 32~words (i.\,e. 64 bytes) of input are loaded from memory into~$W_{\mathrm{in}}$.  Of these words, 31~words are encoded in the iteration with the last word serving as a \emph{look ahead} for surrogate processing (\S~\ref{16to8:surrogates}).  The mask
\begin{equation}
L=1\ll n/2-1 \label{16to8:L}
\end{equation}
indicates the position of the lookahead word in~$W_{\mathrm{in}}$.
\subsection{Classification and Fast Paths}
We first need to find out what UTF-8 cases the characters in our input correspond to.
Comparing the 16-bit words in the input vector with~\texttt{0x0080} and~\texttt{0x0800}, we produce the
masks
\begin{align}
M_{234}&=(\texttt{0x0080}\le W_{\mathrm{in}})\land\lnot L\quad\mbox{and} \label{16to8:M234}\\
M_{12}&=(\texttt{0x0800}> W_{\mathrm{in}}) \label{16to8:M12}
\end{align}
telling us if non-ASCII (i.\,e.~2-, 3-,~or 4-byte) characters and ASCII or 2-byte characters are present.
ASCII characters in the lookahead are ignored to simplify some later bits of the algorithm.
Based on this information, we can then embark on a code path suitable for this chunk of input.

\paragraph{ASCII only} If all input words represent ASCII characters ($M_{234}=0$), we handle the input in an ASCII-only fast path: the vector is truncated to bytes (\texttt{vpmovwb}) and deposited into the output buffer, advancing it by 31~bytes. Though we could advance by 32~bytes,  we want the the algorithm to proceed with a constant stride through memory irrespective of the content.

\paragraph{Default path} If some 3-~or 4-byte characters are present ($M_{12}\lor L\ne\lnot0$), we check for surrogates. We do this by masking the words with~\texttt{0xfc00} and then checking if the result is equal to \texttt{0xd800} (high surrogate,~$M_{\mathrm{hi}}$) or~\texttt{0xdc00} (low surrogate,~$M_{\mathrm{lo}}$), giving
\begin{align}
M_{\mathrm{hi}}&=(\texttt{0xd800}\le W_{\mathrm{in}}<\texttt{0xdc00})\land\lnot L\notag\\
&=(W_{\mathrm{in}}\land\texttt{0xfc00}=\texttt{0xd800})\land\lnot L\quad\mbox{and} \label{16to8:Mhi}\\
M_{\mathrm{lo}}&=(\texttt{0xdc00}\le W_{\mathrm{in}}<\texttt{0xe000})\notag\\
&=(W_{\mathrm{in}}\land\texttt{0xfc00}=\texttt{0xdc00}). \label{16to8:Mlo}
\end{align}
If surrogates are found to be present ($M_{\mathrm{hi}}\lor M_{\mathrm{lo}}\ne0$), we proceed to~\S~\ref{16to8:surrogates} to handle them.\footnote{Low surrogates in the lookahead are registered to permit detection of sequencing errors.}  Otherwise we skip that step, set $W_{\mathrm{joined}}=W_{\mathrm{in}}$ zero-extended from~16-bit to~32-bit (\instruction{vpmovzxwd}), and directly go to~\S~\ref{16to8:encode}.

\paragraph{1/2 byte only} In the third and final case, we know that the input is a mix of ASCII and 2-byte characters.
We process this case by shuffling the bits of two-byte characters into position.\footnote{As we are on a little-endian architecture, the lead byte is the less-significant of the two.}
The most significant two bits of each byte are cleared and tag bits are applied.  Through this whole process, ASCII characters are left unchanged, giving us
\begin{equation} 
W_{\mathrm{out}}=M_{234}\mathrel?(W_{\mathrm{in}}\ll8\lor W_{\mathrm{in}}\gg6)\land\texttt{0x3f3f}\lor\texttt{0x80c0}\mathrel:W_{\mathrm{in}}. \label{16to8:12b:Wout}
\end{equation}
We illustrate this equation in the 2-byte case:
\begin{equation}
\begin{tabular}{rl}
$W_{\mathrm{in}}$&\texttt{0000\,0LKJ\;HGFE\,DCBA}\\
$W_{\mathrm{in}}\ll8\lor W_{\mathrm{in}}\gg6$&\texttt{HGFE\,DCBA\;000L\,KJHG}\\
$W_{\mathrm{out}}$&\texttt{10FE\,DCBA\;110L\,KJHG}\\
\end{tabular}
\end{equation}

The words of $W_{\mathrm{out}}$ before the lookahead are then bytewise compared with \texttt{0x0800}\footnote{a constant we have already loaded into a register; any other constant with high-byte in range \texttt{0x01}--\texttt{0x7f} and low byte~0 works.} producing a mask
\begin{equation}
m_{\mathrm{keep}}=W_{\mathrm{out}}\ge_{\mathrm{byte}}(L\mathrel?\texttt{0xffff}\mathrel:\texttt{0x0800}) \label{16to8:12b:mkeep}
\end{equation}
holding binary~01 for ASCII characters, 11~for 2-byte characters, and 00~for the lookahead.
With this mask, we finally compress~$W_{\mathrm{out}}$ into a UTF-8 stream
\begin{equation}
w_{\mathrm{out}}=\compress(m_{\mathrm{keep}}, W_{\mathrm{out}}) \label{16to8:12b:woutsimpl}
\end{equation}
and write it to the output.

The output buffer pointer is advanced by the number of bytes of output produced, which is one byte for each word of input (sans lookahead) and another byte for each 2-byte character.
\begin{equation}
n_{\mathrm{out}}=\popcount(M_{234})+n/2-1. \label{16to8:12b:nout}
\end{equation}

\subsection{Surrogates} \label{16to8:surrogates}
When surrogates are present in the input, the bits of low surrogate have to be merged into those of the corresponding high surrogate, yielding the code point of the character to be encoded.

First, $W_{\mathrm{in}}$~is zero extended to 32~bits per element.\footnote{From here on, each vector holds $2n$~bytes of data.  These can be implemented as pairs of $n$-byte vectors.}  A vector~$W_{\mathrm{lo}}$, holding for each high surrogate in~$W_{\mathrm{in}}$ its corresponding low surrogate, is produced by rotating~$W_{\mathrm{in}}$ to the right by one element.

Then, the surrogates are joined by subtracting the tag bits (\texttt{0xd800}~for the high surrogate, \texttt{0xdc00} for the low surrogate), undoing the surrogate plane shift for the high surrogate, shifting the bits of the high surrogate into place and then adding the two together.
By pulling out the constants representing the tag bits and the plane shift, these additions and subtractions can be combined into one using 32-bit unsigned arithmetic.
This gives us
\begin{align}
W_{\mathrm{joined}}&=M_{\mathrm{hi}}\mathrel?\bigl((W_{\mathrm{in}}-\texttt{0xd800}+\texttt{0x0040})\ll10\bigr)+(W_{\mathrm{lo}}-\texttt{0xdc00}):W_{\mathrm{in}}\nonumber \\
&=M_{\mathrm{hi}}\mathrel?\bigl((W_{\mathrm{in}}\ll10)-\texttt{0x35f000}\bigr)+(W_{\mathrm{lo}}-\texttt{0xdc00}):W_{\mathrm{in}}\nonumber\\
&=M_{\mathrm{hi}}\mathrel?(W_{\mathrm{in}}\ll10)+W_{\mathrm{lo}}+\texttt{0xfca02400}:W_{\mathrm{in}}. \label{16to8:Wjoined}
\end{align}

With the surrogate pairs decoded, we can then proceed to~\S~\ref{16to8:encode} to encode into UTF-8.  The vector elements corresponding to low surrogates are ignored for the rest of the algorithm.

\subsection{Encoding into UTF-8} \label{16to8:encode}
When we reach this step, we have transformed~$W_{\mathrm{in}}$ into a vector~$W_{\mathrm{joined}}$ of 32-bit integers, holding the code points of the characters in the input.\footnote{In the presence of surrogates, some of these elements are ignored.}  We would now like to encode these code points into UTF-8, producing 1--4~bytes of output per code point.

Consider Fig.~\ref{16to8:correspondence}: for the 2-, 3-~and 4-byte case, the bits~\texttt A--\texttt W making up the code point always appear in the same position.
This suggests using the same encoding procedure for the 2-, 3-,~and 4-byte case with merely different tag bits applied at the end.
ASCII characters are handled with a  shift into position.

The encoding procedure is based on the \texttt{vpmultishiftqb} instruction introduced with the VBMI instruction set extension.  Given a vector of 64-bit words and for each such word a vector of eight bytes, the instruction uses the byte vectors as indices to pick eight 8-bit chunks of data (8~consecutive bits) from the corresponding source words.
By choosing these indices such that they do not cross a 32-bit boundary, we can effectively use the instruction to select four 8-bit chunks out of each 32-bit word.

Applying the index vector~$(18,12,6,0)$ to each 32-bit word\footnote{i.\,e.\ the index vector~$(18,12,6,0,50,44,38,32)$ applied to each 64-bit word} of~$W_{\mathrm{joined}}$, we obtain~$W_{\mathrm{shifted}}$ with each bit shifted into the right position with some bits left over:
\begin{equation} \label{16to8:Wshifted}
\begin{tabular}{rl}
$W_{\mathrm{joined}}$&\texttt{0000\,0000\;000w\,vuts\;RQPN\,MLKJ\;HGFE\,DCBA}\\
\midrule
index          18&\texttt{\hphantom{0000\,00}00\;000w\,vu}\\
index          12&\texttt{\hphantom{0000\,0000\;0000\,}vuts\;RQPN\,    \;    \,    }\\
index \hphantom06&\texttt{\hphantom{0000\,0000\;0000\,0000\;00}PN\,MLKJ\;HG  \,    }\\
index \hphantom00&\texttt{\hphantom{0000\,0000\;0000\,0000\;0000\,0000\;}HGFE\,DCBA}\\
\midrule
$W_{\mathrm{shifted}}$&\texttt{HGFE\,DCBA\;PNML\,KJHG\;vuts\,RQPN\;0000\,0wvu}\\
\end{tabular}
\end{equation}
To fix up the left-over bits, we mask with~\texttt{0x3f3f3f3f}, reusing the mask from the 2-byte fast path.
Then, appropriate tag bits~$W_{\mathrm{tag}}$ are applied:
\begin{equation} \label{16to8:Wtag}
\begin{tabular}{ccc}
\emph{case}&\emph{$W_{\mathrm{shifted}}$ masked with \texttt{\upshape0x3f3f3f3f}}&\emph{tag bits}\\
\midrule
2-byte&\texttt{00FE\,DCBA\;000L\,KJHG\;0000\,0000\;0000\,0000}&\texttt{0x80c00000}\\
3-byte&\texttt{00FE\,DCBA\;00ML\,KJHG\;0000\,RQPN\;0000\,0000}&\texttt{0x8080e000}\\
4-byte&\texttt{00FE\,DCBA\;00ML\,KJHG\;00ts\,RQPN\;0000\,0wvu}&\texttt{0x808080f0}\\
\end{tabular}
\end{equation}

Finally, the ASCII case is handled by just shifting the ASCII words into position and
merging these shifted characters into the output of the other cases, giving us
\begin{equation}
W_{\mathrm{out}}=M_{234}\mathrel?W_{\mathrm{shifted}}\land\texttt{0x3f3f3f3f}\lor W_{\mathrm{tag}}\mathrel:W_{\mathrm{in}}\ll24. \label{16to8:Wout}
\end{equation}

We end up with UTF-8 encoded characters in~$W_{\mathrm{out}}$.  Each character occupies a 32-bit word and is padded with~\texttt{0x00} bytes to 4~bytes.
Input words corresponding to low surrogates have been passed through, being decoded into junk content.
We get rid of the padding and the low surrogate junk by preparing a mask of bytes we want to keep and compressing out the unwanted bytes using the \texttt{vpcompressb} instruction.

In the mask, we want to keep the most significant byte of each 32-bit word and all non-zero bytes---except for processed low surrogates.
These seemingly complex requirements can be negotiated in two steps by first building a comparison mask and then taking all bytes that are not lower than the mask.
For low surrogate bytes and the lookahead, the mask is~\texttt{0xff} which cannot occur in~$w_{\mathrm{out}}$.\footnote{The byte value \texttt{0xff} is not possible in valid UTF-8 (cf.~Table~\ref{table:tagbits}).  It could be generated through invalid sequencing of surrogates, but we catch such an error when validating the content.}
For the most significant byte of all other words, it is~\texttt{0x00} which admits every byte.
For other bytes, it is~\texttt{0x01}, admitting only nonzero bytes.
Thus we have
\begin{align}
W_{\mathrm{keep}}&=M_{\mathrm{lo}}\lor L\mathrel?\texttt{0xffffffff}\mathrel:\texttt{0x00010101},\quad\mbox{building} \label{16to8:Wkeep}\\
m_{\mathrm{keep}}&=(W_{\mathrm{out}}\ge_{\mathrm{byte}}W_{\mathrm{keep}}) \label{16to8:mkeep}.
\end{align}

With this mask, we compress~$W_{\mathrm{out}}$ into
\begin{equation}
w_{\mathrm{out}}=\compress(m_{\mathrm{keep}}, W_{\mathrm{out}}), \label{16to8:wout}
\end{equation}
write it to the output buffer and advance the output by
\begin{equation}
n_{\mathrm{out}}=\popcount(m_{\mathrm{keep}}) \label{16to8:nout}
\end{equation}
bytes.
Due to the little-endian orientation of the x64 architecture, the bytes of each UTF-8 sequence end up in the right order: within each 32-bit word, they are written from the least significant byte to the most significant byte.

\subsection{Validation}
In contrast to the UTF-8 to UTF-16 procedure, validation of UTF-16 input is less involved.
We merely have to check for the correct sequencing of surrogates: every high surrogate must be followed by a low surrogate and vice versa.
As this validation only pertains surrogates, it is skipped in their absence, i.\,e.~in all fast paths;
input strings without surrogates are always valid.

To aid in this process, we only process 31~words of input in each iteration, permitting a ``look ahead'' into the first word of the next iteration.
We also keep track of a \emph{surrogate carry~$c$} indicating if the first word in~$W_{\mathrm{in}}$ was preceded by a high surrogate.
This carry allows us to decide if a low surrogate in~$W[0]$ is to be ignored~($c=1$) or is a sequencing error~($c=0$).\footnote{When~$W[0]$ is not a low surrogate, $c$~is guaranteed to be clear.}

Correct sequencing is checked for by concatenating $M_{\mathrm{hi}}$ with~$c$ and shifting it to the position of the corresponding low surrogates~$M_{\mathrm{lo}}$.
The input is valid if each high surrogate corresponds to a low surrogate:
\begin{equation} \label{16to8:valsur}
\mbox{valid if}\quad (M_{\mathrm{hi}}\ll1\lor c)=M_{\mathrm{lo}}.
\end{equation}

The carry for the next iteration is computed as the presence of a high surrogate in the vector element right before the lookahead, giving
\begin{equation}
c_{\mathrm{out}}=M_{\mathrm{hi}}\gg(n/2-2)\land1. \label{16to8:cout}
\end{equation}
In the absence of surrogates, i.\,e.~in the fast paths, the carry is  cleared ($c_{\mathrm{out}}=0$).

If validation fails, we find the location of the first mismatched surrogate to transcode the words preceding the encoding error and then terminate.
This is done by setting the number of remaining input words~$\ell$ to the number of words preceding the encoding error and then jumping to the tail handling code~\S~\ref{sec:tail}.

Computing the location requires more work than Eq.~\ref{16to8:valsur}; for a high surrogate not followed by a low surrogate, that equation indicates the missing low surrogate as the first erroneous word when it should really be the unmatched high surrogate.
So we proceed more carefully and first compute the sets of high surrogates not followed by low surrogates
\begin{equation}
M_{\mathrm{hi-lo}}=M_{\mathrm{hi}}\land\lnot(M_{\mathrm{lo}}\gg1) \label{16to8:Mhilo}
\end{equation}
and the set of low surrogates not preceded by high surrogates
\begin{equation}
M_{\mathrm{lo-hi}}=M_{\mathrm{lo}}\land\lnot(M_{\mathrm{hi}}\ll1\lor c). \label{16to8:Mlohi}
\end{equation}
The number of valid bytes is then the longest prefix not found in either of these masks:
\begin{equation}
\ell=\ctz(M_{\mathrm{lo-hi}}\lor M_{\mathrm{hi-lo}}). \label{16to8:l}
\end{equation}

\subsection{Decoding Failure and Tail Handling} \label{sec:tail}

At the end of the input, there might be some UTF-16 words left to process, but not enough to load a whole 64-byte vector.
We deal with this remaining input in a manner similar to the UTF-8 to UTF-16 case,~cf.~\S~\ref{8to16:tail}.

When $\ell<n/2$~words of input remain to be processed, we compute a mask
\begin{equation}
B=(1\ll\ell)-1 \label{16to8:B}
\end{equation}
of input words left to be processed.
We then load the remaining input zero-masked with~$B$,\footnote{Or in case of an encoding
error, mask the already loaded vector.} giving us the input tail padded with~\codepoint{0000}.
This remaining input is then processed in a final iteration of the main loop.
As each null~byte translates into a single byte of output, this leads to an output that is precisely $n/2-1-\ell$~bytes longer than the true output length.
We compensate for this by adjusting the output length reported to the caller accordingly.

In contrast to the other direction, this approach may write past the end of the output buffer if it is just long enough to hold the decoded string.
We avoid this problem by performing masked stores instead of potentially storing null~bytes past the end of the output.

\section{Experiments}\label{sec:experiments}
Our initial implementation of the algorithms was written in Intel~64 assembly for systems following the System V ABI~\cite{sysvabi}.
Hand-tuned assembly can slightly surpass optimizing
compilers due to better instruction scheduling and register allocation.
For the measurements and comparison with competitive libraries, we have translated the code to C++ using \emph{intrinsic functions}~\cite{intelintrinsic} to access AVX-512 instructions and integrated it into our \emph{simdutf} library\footnote{\url{https://github.com/simdutf/simdutf}} as the \emph{AVX-512} kernel.
This library is freely available.
Despite a slight loss of performance in comparison to the assembly implementation, we believe that this approach facilitates better portability and integration into existing software.

Our library is organized in different kernels that are automatically selected at runtime based on the features of the CPU, a process sometimes called runtime dispatching.
During benchmarking, we can manually select the different kernels.
As the names suggest,  the AVX2 kernel relies on AVX2 instructions (32-byte vector length) while the AVX-512 kernel using our new functions relies on AVX-512 instructions with a 64-byte vector length.
Our new functions are part of the AVX-512 kernel, and the AVX2 kernel represents results presented by Lemire and Mu{\l}a~\cite{lemire2022transcoding}.

For benchmarking, we use Ubuntu~22.04 on a non-virtual (\emph{metal}) server from Amazon Web Services (\texttt{c6i.metal}).
These servers have 32-core Intel Xeon 8375C (Ice Lake) processors with \SI{41}{\mebi\byte} of L3 memory, with \SI{48}{\kilo\byte} of L1~data cache memory and \SI{1.25}{\mebi\byte} of L2 cache memory per core.
The base clock frequency is \SI{2.9}{\giga\hertz}, with a maximal frequency of \SI{3.5}{\giga\hertz}.
They have \SI{256}{\gibi\byte} of main memory (DDR4, \SI{3200}{\mega\hertz}).
The benchmarks are single-threaded and we exclude disk and network accesses from our tests.
The software is written in C++ and compiled with the Clang~14 C++ compiler from the LLVM project using the default \texttt{cmake} setting for a release build: \texttt{-O3 -DNDEBUG}. 

We could use several threads. 
For example, we could split the input into segments,
and compute the expected transcoded size
of the segments, before transcoding each segment in
its own thread.
However, merely joining a thread under Linux can require tens of microseconds of waiting from the main thread. With the high speed of our functions,
this penalty is equivalent to the time required to process hundreds of kilobytes of data. We could
use faster synchronization techniques (e.g., spin~locks and thread pools), but at the expense of complexity and power efficiency. We expect
that multicore parallelism is only warranted for large inputs, in the megabytes or gigabytes range. Future work might consider such cases.

\subsection{Setup}
We benchmark the transcoding of data files between UTF-8 and UTF-16 in memory. 
We repeat the task \num{10000}~times, measuring the time of each conversion: the C++ library reports a precision of \SI{1}{\nano\second} for the \texttt{std::chrono::steady\_clock} measures on our test system~\cite{10.1145/3458336.3465293}. 
The distribution of timings has a long tail akin to a log-normal distribution: 
most values are close to the minimum.
We verify automatically that the difference between the minimum and the average timing is small (less than $1\,\%$).

AVX-512 capable Intel processors prior to the Ice Lake and Rocket Lake families 
would systematically reduce their frequency when using 512-bit instructions, 
a process that Intel referred to as \emph{licensing}. Such 512-bit licensing 
is no longer present in the more recent processors~\cite{travisdownclocking}. 
However,  the processor frequency may fluctuate based on power consumption 
and heat production as is generally the case with Intel processors. We expect 
512-bit instructions to use more power, and thus to run at a slightly lower frequency. 
Irrespective of power usage,  Intel processors execute 512-bit instructions at a 
reduced speed  initially (e.g., $4\times$ slower)---for a few microseconds. 
We assume that our functions with 512-bit instructions are part of a binary executable 
compiled with optimizations for 512-bit capable processors so that this temporary 
effect is uncommon, maybe occurring only once.

We are interested in the steady-state performance of our functions: we
therefore always  benchmark our functions twice: once to intuitively \emph{warm} 
the processor so that 512-bit instructions always execute at full speed and 
so that the processor has had a chance to decode the instructions. 
Furthermore, we may sometimes benchmark a function relying on 512-bit instructions, 
followed by a conventional function: to ensure that the latter is not penalized 
by the power usage of the first function, we pause for a millisecond when switching
the benchmarked function.

We report performance results in characters per second. A given string has the same 
number of characters irrespective of the format (UTF-8, UTF-16). We also report speeds in gigabytes per second by taking the size of the input and dividing by the time elapsed. We focus on little-endian 
UTF-16, but our software supports big-endian UTF-16, at little cost.

We compare our work with the following competitors:
\begin{itemize}
\item We use the \texttt{u8u16} library~\cite{cameron2008case} (last released in 2007).
\item We use the \texttt{utf8lut} library~\cite{stgatilov} (last modified April~19, 2020).
We require full validation of input (\texttt{cmValidate}).
\item  We use the C++ component from International Components for Unicode (ICU)~\cite{icu}. 
\item We use the transcoding functions of the LLVM project, they were originally produced by  the Unicode Consortium.
\item We use the \texttt{iconv} library, which is part of the C library (GNU~C Library~2.35).
\end{itemize}

We use automatically generated (lipsum) text in Arabic, Chinese, Hebrew, Hindi, Japanese, Korean, Latin
and Russian, as well as a list of emojis (henceforth Emoji).\footnote{\url{https://github.com/rusticstuff/simdutf8}}
When formatted as UTF-16, they range in size between \SI{11}{\kibi\byte} and
\SI{170}{\kibi\byte}.
The Chinese, Hindi, Japanese and Korean files have a high fraction of 3-byte UTF-8 characters.
The Arabic, Hebrew and Russian files have a high fraction of 2-byte UTF-8 characters.
Except for the Emoji file, none of the file contain 4-byte UTF-8 characters.
We  make our files freely available.\footnote{\url{https://github.com/lemire/unicode_lipsum}}

\begin{figure}\centering
\subfloat[UTF-8 to UTF-16 transcoding]{
\includegraphics[width=0.49\textwidth]{lipsumspeedutf8utf16.tikz}
}
\subfloat[UTF-16 to UTF-8 transcoding]{
\includegraphics[width=0.49\textwidth]{lipsumspeedutf16utf8.tikz}
}

\caption{\label{fig:transcoding} Transcoding speeds in gigabytes of input data per second for various test files. We compare against the ICU library.
The simdutf library provides both  AVX2 and AVX-512  functions.}
\end{figure}

\begin{table}[tbh!]\centering
\caption{\label{table:lipsumutf8utf16} Validating transcoding speeds (gigacharacters per second) over the lipsum datasets, last row is the harmonic mean of the column.  The last column (AVX-512) presents the results from our new algorithms.}
\small
\subfloat[UTF-8 to UTF-16]{\restartrowcolors
\begin{tabular}{lggggggg}\toprule
   & \mc{llvm} & \mc{iconv} & \mc{ICU} & \mc{u8u16} & \mc{utf8lut} & \mc{AVX2} & \mc{AVX-512}\\\midrule
Arabic & 0.17 & 0.39 & 0.80 & 0.87 & 0.92 & 1.3 & 4.3 \\
Chinese & 0.22 & 0.26 & 0.50 & 0.45 & 0.63 & 1.2 & 1.8 \\
Emoji & 0.18 & 0.19 & 0.22 & 0.31 & 0.18 & 0.40 & 1.0 \\
Hebrew & 0.17 & 0.39 & 0.80 & 0.87 & 0.92 & 1.3 & 4.3 \\
Hindi & 0.16 & 0.21 & 0.43 & 0.49 & 0.72 & 0.84 & 1.7 \\
Japanese & 0.21 & 0.26 & 0.51 & 0.46 & 0.64 & 1.2 & 1.7 \\
Korean & 0.13 & 0.30 & 0.62 & 0.54 & 0.72 & 0.89 & 1.8 \\
Latin & 0.35 & 0.56 & 1.5 & 13. & 1.0 & 22. & 20. \\
Russian & 0.17 & 0.29 & 0.46 & 0.86 & 0.92 & 1.3 & 4.2 \\
harm.\ mean  &0.18 &0.29 &0.50 &0.60 &0.57 &1.0 &2.3 \\
\bottomrule
\end{tabular}
}\\
\subfloat[UTF-16 to UTF-8]{\restartrowcolors
\begin{tabular}{lggggggg}\toprule
   & \mc{llvm} & \mc{iconv} & \mc{ICU} & \mc{utf8lut} & \mc{AVX2} & \mc{AVX-512} \\\midrule
Arabic & 0.38 & 0.30 & 0.67 & 2.4 & 4.8 & 11. \\
Chinese & 0.38 & 0.28 & 0.36 & 2.4 & 2.6 & 3.9 \\
Emoji & 0.29 & 0.20 & 0.27 & 0.37 & 0.38 & 1.6 \\
Hebrew & 0.48 & 0.32 & 0.68 & 2.3 & 4.8 & 11. \\
Hindi & 0.31 & 0.21 & 0.21 & 2.4 & 2.6 & 3.8 \\
Japanese & 0.38 & 0.26 & 0.37 & 2.3 & 2.7 & 3.8 \\
Korean & 0.43 & 0.30 & 0.37 & 2.3 & 2.7 & 3.8 \\
Latin & 0.58 & 0.56 & 0.91 & 2.3 & 18. & 20. \\
Russian & 0.28 & 0.23 & 0.23 & 2.4 & 4.8 & 11. \\
harm.\ mean  &0.37 &0.27 &0.36 &1.5 &1.9 &4.5 \\
\bottomrule
\end{tabular}
}
\end{table}

\begin{table}[tbh!]\centering
\caption{\label{table:lipsumutf8utf16gbs} Validating transcoding speeds in gigabytes of input per second over the lipsum datasets, last row is the harmonic mean of the column. The last column (AVX-512) presents the results from our new algorithms.}
\small
\subfloat[UTF-8 to UTF-16]{\restartrowcolors
\begin{tabular}{lggggggg}\toprule
   & \mc{llvm} & \mc{iconv} & \mc{ICU} & \mc{u8u16} & \mc{utf8lut} & \mc{AVX2} & \mc{AVX-512}\\\midrule
Arabic & 0.31 & 0.70 & 1.4 & 1.5 & 1.6 & 2.3 & 7.6 \\
Chinese & 0.66 & 0.77 & 1.5 & 1.3 & 1.9 & 3.7 & 5.4 \\
Emoji & 0.72 & 0.78 & 0.87 & 1.2 & 0.70 & 1.6 & 4.2 \\
Hebrew & 0.31 & 0.70 & 1.4 & 1.6 & 1.6 & 2.3 & 7.7 \\
Hindi & 0.43 & 0.56 & 1.2 & 1.3 & 1.9 & 2.3 & 4.6 \\
Japanese & 0.60 & 0.74 & 1.5 & 1.3 & 1.9 & 3.5 & 5.0 \\
Korean & 0.32 & 0.73 & 1.5 & 1.3 & 1.8 & 2.2 & 4.5 \\
Latin & 0.35 & 0.56 & 1.5 & 13. & 1.0 & 22. & 20. \\
Russian & 0.31 & 0.52 & 0.83 & 1.6 & 1.7 & 2.4 & 7.7 \\
harm.\ mean  &0.40 &0.66 &1.2 &1.5 &1.4 &2.6 &6.0 \\
\bottomrule
\end{tabular}
}\\
\subfloat[UTF-16 to UTF-8]{\restartrowcolors
\begin{tabular}{lggggggg}\toprule
   & \mc{llvm} & \mc{iconv} & \mc{ICU} & \mc{utf8lut} & \mc{AVX2} & \mc{AVX-512} \\\midrule
Arabic & 0.76 & 0.61 & 1.3 & 4.7 & 9.6 & 21. \\
Chinese & 0.76 & 0.57 & 0.73 & 4.7 & 5.3 & 7.7 \\
Emoji & 1.2 & 0.82 & 1.1 & 1.5 & 1.5 & 6.5 \\
Hebrew & 0.96 & 0.63 & 1.3 & 4.7 & 9.6 & 21. \\
Hindi & 0.63 & 0.42 & 0.41 & 4.7 & 5.3 & 7.7 \\
Japanese & 0.77 & 0.52 & 0.74 & 4.7 & 5.3 & 7.6 \\
Korean & 0.86 & 0.60 & 0.73 & 4.7 & 5.4 & 7.6 \\
Latin & 1.2 & 1.1 & 1.8 & 4.7 & 37. & 40. \\
Russian & 0.56 & 0.45 & 0.46 & 4.7 & 9.6 & 21. \\
harm.\ mean  &0.80 &0.59 &0.77 &3.8 &5.1 &11. \\
\bottomrule
\end{tabular}
}
\end{table}

\subsection{Results}

We present speed results in gigacharacters per second regarding the validating UTF-8 to UTF-16 transcoding functions on the lipsum files in Table~\ref{table:lipsumutf8utf16}.
The AVX2 and AVX-512 columns correspond to our own code, in the simdutf library (version~2.0.9, git tag \texttt{v2.0.9}).
As the AVX2 kernels are implemented with a 32-byte vector length compared to the 64-byte vector length of the AVX-512 kernels, we expect a $33\,\%$ higher throughput just from using longer vectors (see~\S~\ref{avx512:perf}).

On the Latin file, the new AVX-512 kernel fails to improve on the earlier AVX2 kernel: the Latin dataset is made almost entirely of ASCII inputs which the AVX-512 kernel processes 32~bytes at a time, just like the AVX2 kernel.
On the Emoji file, the new AVX-512 kernel achieves one~gigacharacter per second when transcoding from UTF-8, which is more than twice as fast as any competitor.
When transcoding from UTF-16, the new kernel transcode the Emoji file at 1.6~gigacharacters per second, which is more than four times as fast as any competitor. Whether transcoding from UTF-8 or UTF-16, the new kernel does well when transcoding inputs dominated by 2-byte UTF-8 sequences (Arabic, Hebrew and Russian): it is twice as fast as any competitor. For inputs dominated by 3-byte UTF-8 sequences (Chinese, Hindi, Japanese and Korean), the gain compared to the earlier AVX2 kernel is of the order of 50\%.

Fig.~\ref{fig:transcoding} and Table~\ref{table:lipsumutf8utf16gbs} present the speed results in gigabytes of inputs per second.
Whereas the speeds of the AVX-512 function in gigacharacters per second vary by multiples (from 4.3 with Arabic to 1.0 with Emoji), the gaps are much less significant in gigabytes per second (from 7.6 with Arabic to 4.2 with Emoji). 

Table~\ref{table:lipsumutf8utf16ins} presents the number of instructions retired per character,
measured using the hardware performance counters provided by Intel.\footnote{Under Linux, the performance counters are made available by the operating system. A C~program can query them using the
functions defined in the \texttt{linux/perf\_event.h}~header.}
In the worst case (for the Emoji files), the new AVX-512 kernel still requires fewer than 6~instructions per character to transcode in either direction.
Except for  the Latin files, the new AVX-512 kernel requires far fewer than half the number of instructions than the AVX2 kernel when transcoding from UTF-8.
For example, we reduce the number of instructions by a factor of three for the Arabic file.
Table~\ref{table:lipsumutf8utf16insc} provides the number of instructions per cycle. We find that the AVX-512 kernel is associated with a lower number of instructions retired per cycle---especially so when transcoding from UTF-8. Correspondingly, we expect a lower number of 64-byte instructions being retired per cycle compared to 32-byte instructions due to the microarchitectures of the Intel CPUs (\S~\ref{avx512:perf}).

The \texttt{utf8lut} library, when transcoding from UTF-8, requires fewer instructions than our AVX-2 kernel, but it is associated with few instructions per cycle. Hence, the \texttt{utf8lut} library is generally slower than our AVX-2 kernel despite relying on the same instruction set. The \texttt{utf8lut} library relies on a \SI{2}{\mebi\byte} table for UTF-8 to UTF-16 transcoding as opposed to a small table (\SI{11}{\kibi\byte}) for our AVX-2 kernel, and no table at all for our AVX-512 kernel. A large table may cause the CPU to wait for loads to complete and increases overall cache pressure.

\begin{table}[tbh!]\centering
\caption{\label{table:lipsumutf8utf16ins}
CPU instructions retired per character when transcoding with validation.  The last column (AVX-512) presents the results from our new algorithms.}
\small
\subfloat[UTF-8 to UTF-16]{\restartrowcolors
\begin{tabular}{lggggggg}\toprule
   & \mc{llvm} & \mc{iconv} & \mc{ICU} & \mc{u8u16} & \mc{utf8lut} & \mc{AVX2} & \mc{AVX-512} \\\midrule
Arabic & 65. & 52. & 27. & 15. & 5.3 & 7.4 & 2.3 \\
Chinese & 82. & 78. & 38. & 31. & 8.4 & 11. & 3.2 \\
Emoji & 100 & 100 & 93. & 45. & 95. & 29. & 5.4 \\
Hebrew & 65. & 51. & 27. & 15. & 5.3 & 7.4 & 2.3 \\
Hindi & 80. & 71. & 34. & 28. & 7.3 & 12. & 3.0 \\
Japanese & 81. & 76. & 37. & 30. & 8.2 & 11. & 3.2 \\
Korean & 75. & 66. & 31. & 25. & 6.9 & 12. & 2.7 \\
Latin & 56. & 35. & 11. & 0.65 & 5.3 & 0.35 & 0.24 \\
Russian & 66. & 52. & 27. & 16. & 5.3 & 7.2 & 2.3 \\
\bottomrule
\end{tabular}
}\\
\subfloat[UTF-16 to UTF-8]{\restartrowcolors
\begin{tabular}{lggggggg}\toprule
   & \mc{llvm} & \mc{iconv} & \mc{ICU} & \mc{utf8lut} & \mc{AVX2} & \mc{AVX-512} \\\midrule
Arabic & 37. & 53. & 27. & 6.3 & 2.6 & 1.1 \\
Chinese & 48. & 67. & 41. & 6.3 & 4.5 & 2.2 \\
Emoji & 62. & 90. & 67. & 51. & 48. & 5.4 \\
Hebrew & 37. & 53. & 27. & 6.3 & 2.6 & 1.1 \\
Hindi & 45. & 62. & 38. & 6.3 & 4.5 & 2.2 \\
Japanese & 47. & 65. & 40. & 6.3 & 4.5 & 2.2 \\
Korean & 43. & 58. & 35. & 6.3 & 4.5 & 2.2 \\
Latin & 31. & 34. & 19. & 6.3 & 0.69 & 0.55 \\
Russian & 37. & 53. & 27. & 6.3 & 2.6 & 1.1 \\
\bottomrule
\end{tabular}
}
\end{table}

\begin{table}[tbh!]\centering
\caption{\label{table:lipsumutf8utf16insc}
CPU instructions retired per cycle when transcoding with validation.  The last column (AVX-512) presents the results from our new algorithms.}
\small
\subfloat[UTF-8 to UTF-16]{\restartrowcolors
\begin{tabular}{lggggggg}\toprule
   & \mc{llvm} & \mc{iconv} & \mc{ICU} & \mc{u8u16} & \mc{utf8lut} & \mc{AVX2} & \mc{AVX-512} \\\midrule
Arabic & 3.3 & 5.8 & 6.1 & 3.8 & 1.4 & 2.7 & 2.9 \\
Chinese & 5.2 & 5.7 & 5.3 & 4.0 & 1.5 & 3.9 & 1.7 \\
Emoji & 5.1 & 5.8 & 5.8 & 3.9 & 4.8 & 3.3 & 1.6 \\
Hebrew & 3.2 & 5.8 & 6.1 & 3.8 & 1.4 & 2.7 & 2.9 \\
Hindi & 3.7 & 4.3 & 4.2 & 3.9 & 1.5 & 2.9 & 1.5 \\
Japanese & 4.8 & 5.5 & 5.4 & 3.9 & 1.5 & 3.9 & 1.6 \\
Korean & 2.8 & 5.6 & 5.4 & 3.9 & 1.4 & 3.1 & 1.5 \\
Latin & 5.6 & 5.6 & 4.6 & 2.4 & 1.6 & 2.2 & 1.4 \\
Russian & 3.2 & 4.3 & 3.6 & 3.9 & 1.4 & 2.7 & 2.9 \\
harm.\ mean   &3.9 &5.3 &5.0 &3.6 &1.6 &3.0 &1.8 \\
\bottomrule
\end{tabular}
}\\
\subfloat[UTF-16 to UTF-8]{\restartrowcolors
\begin{tabular}{lggggggg}\toprule
   & \mc{llvm} & \mc{iconv} & \mc{ICU} & \mc{utf8lut} & \mc{AVX2} & \mc{AVX-512} \\\midrule
Arabic & 4.0 & 4.6 & 5.2 & 4.2 & 3.5 & 3.3 \\
Chinese & 5.2 & 5.4 & 4.8 & 4.2 & 3.4 & 2.5 \\
Emoji & 5.2 & 5.3 & 5.2 & 5.3 & 5.1 & 2.6 \\
Hebrew & 5.1 & 4.8 & 5.2 & 4.2 & 3.5 & 3.3 \\
Hindi & 4.1 & 3.7 & 2.6 & 4.2 & 3.4 & 2.5 \\
Japanese & 5.2 & 4.9 & 4.7 & 4.2 & 3.4 & 2.5 \\
Korean & 5.3 & 5.0 & 4.5 & 4.2 & 3.4 & 2.5 \\
Latin & 5.2 & 5.5 & 5.0 & 4.2 & 3.6 & 3.2 \\
Russian & 3.0 & 3.4 & 2.1 & 4.2 & 3.5 & 3.3 \\
harm.\ mean  &4.5 &4.6 &3.9 &4.3 &3.6 &2.8 \\
\bottomrule
\end{tabular}
}
\end{table}

In Fig.~\ref{fig:transcodingvariouslengths}, we present the measured transcoding speed for various small prefixes of the Arabic files. We find that as long as the input has hundreds of characters, we can reach and exceed a billion characters decoded per second.

\begin{figure}[tbh!]\centering
\begin{tikzpicture}
\begin{axis}[
      grid=none,
      mark = none,
      xmin=35,
      xmax=3000,
      axis x line*=bottom,
      axis y line*=left,
      ylabel={billions of characters per second},
	  xlabel={number of characters (prefix)},
	  legend pos=south east]
\addplot[black!30!blue, ultra thick] table [x=n, y=speed, col sep=comma] {arabicutf16.dat};
\addlegendentry{UTF-16 to UTF-8};
\addplot[black!30!red, ultra thick] table [x=n, y=speed, col sep=comma] {arabicutf8.dat};
\addlegendentry{UTF-8 to UTF-16};
\end{axis}
\end{tikzpicture}

\caption{\label{fig:transcodingvariouslengths} Validating transcoding speed in billions of characters per second for prefixes of various lengths of the Arabic files using our techniques. }
\end{figure}

Historically, some processors could only read and write data when the memory address was a multiple of the data size. Older Intel processors could read and write at any address, but with a severe penalty for  unaligned memory addresses.
On recent processors (e.g., Intel's  Sandy Bridge microarchitecture launched in 2011), there is reportedly no measurable performance penalty for reading or writing misaligned memory operands~\cite{agner3}. However, there might be indirect penalties (e.g., accessing more cache lines).
In the hope of achieving better performance, we could require that our memory buffers start at an address divisible by 512~bits. However, we expect that the performance of the transcoding functions is generally unaffected by memory alignment on our test system.
Thus our benchmarking code does not align the memory in any particular manner, relying instead on the default behavior of the memory allocator. 
To test the effect of the memory alignment of the input, we  transcoded the same data, but shifted by 0~to 512~bytes inside a buffer. Using one of our UTF-8 file (Arabic), we measured a difference of 2\% between the fastest and slowest alignment when using our fast (AVX-512) transcoder. 
We get a similar result if we transcode from a fixed input to offsetted locations inside a destination buffer. Our results suggest that memory alignment is likely not a significant factor.
\section{Conclusion}

It is not \emph{a priori} obvious that character transcoding is amenable
to SIMD processing. %
Earlier work achieved
high speeds but it required kilobytes of lookup tables~\cite{lemire2022transcoding}. 
Our work indicates that the AVX-512 instruction-set extensions enables high speed for
tasks such as character transcoding---without lookup tables and using few instructions. 
It suggests that some features of the AVX-512 instruction-set extensions 
might serve as a reference for future instruction-set extensions. In particular, we find masked SIMD instructions (move, load, store, compress) with byte-level granularity useful.

Both Intel and AMD support AVX-512 instructions. They also both offer specialized compilers, tuned for their processors. Future work could compare the performance of our routines on more varied Intel and AMD  processors (e.g., Intel Rocket Lake and Sapphire Rapids, AMD Zen~4), using specialized compilers (e.g., from Intel and AMD) and hand-tuned assembly.
We could extend our benchmarks to cover a wider range of string.

\section*{Acknowledgements}

We thank W.~Mu\l a who produced an early UTF-8 to UTF-16 transcoder using AVX-512
instructions. The version presented in this manuscript follows a different
design but Mu\l a's work provided a crucial motivation.
We thank N.~Boyer for his technical work on our software library, benchmarks,
and tests. 

\bibliography{avx512-utf}
\clearpage
\appendix
\section{UTF-8 to UTF-16: Summary of Variables}
\label{appendix:utf8utf16}
\begin{equation*}
\begin{tabular}{ccp{0.535\textwidth}c}
\toprule
\emph{symbol}&\emph{type}&\emph{description}&\emph{Eq.}\\
\midrule
$w_{\mathrm{in}}$&byte vector&input vector&---\\
$m_1$&byte mask&1-byte sequence lead bytes in~$w_{\mathrm{in}}$&(\ref{8to16:m1})\\
$m_{234}$&byte mask&2/3/4-byte sequence lead bytes in~$w_{\mathrm{in}}$&(\ref{8to16:m234})\\
$m_2$&byte mask&2-byte sequence lead bytes in~$w_{\mathrm{in}}$&---\\
$m_{34}$&byte mask&3/4-byte sequence lead bytes in~$w_{\mathrm{in}}$&(\ref{8to16:m34})\\
$m_3$&byte mask&3-byte sequence lead bytes in~$w_{\mathrm{in}}$&(\ref{8to16:m3})\\
$m_4$&byte mask&4-byte sequence lead bytes in~$w_{\mathrm{in}}$&(\ref{8to16:m4})\\
$m_{1234}$&byte mask&lead bytes in~$w_{\mathrm{in}}$&(\ref{8to16:m1234})\\
$m_{+1}$&byte mask&second byte of each sequence in~$w_{\mathrm{in}}$&(\ref{8to16:mplus1})\\
$m_{+2}$&byte mask&third byte of each sequence in~$w_{\mathrm{in}}$&(\ref{8to16:mplus2})\\
$m_{+3}$&byte mask&fourth byte of each sequence in~$w_{\mathrm{in}}$&(\ref{8to16:mplus3})\\
$m_{\mathrm{end}}$&byte mask&last byte of each sequence in~$w_{\mathrm{in}}$&(\ref{8to16:mend})\\
$M_3$&word mask&3-byte characters in~$W_{\mathrm{out}}$&(\ref{8to16:M3})\\
$M_{\mathrm{hi}}$&word mask&high surrogates in~$W_{\mathrm{out}}$&(\ref{8to16:Mhi})\\
$M_{\mathrm{lo}}$&word mask&low surrogates in~$W_{\mathrm{out}}$&(\ref{8to16:Mlo})\\
$w_{\mathrm{stripped}}$&byte vector&$w_{\mathrm{in}}$ with tag bits stripped off&(\ref{8to16:wstripped})\\
$P$&word vector&indices of last-in-sequence bytes in~$w_{\mathrm{in}}$&(\ref{8to16:P})\\
$m_{-1}$&byte mask&mask to admit only $2^{\mathrm{nd}}$-last bytes of~$w_{\mathrm{in}}$&(\ref{8to16:mminus1})\\
$m_{-2}$&byte mask&mask to admit only $3^{\mathrm{rd}}$-last bytes of~$w_{\mathrm{in}}$&(\ref{8to16:mminus2})\\
$W_{\mathrm{end}}$&word vector&last byte of each sequence in~$w_{\mathrm{in}}$&(\ref{8to16:Wend})\\
$W_{-1}$&word vector&second-last byte of each sequence in~$w_{\mathrm{in}}$&(\ref{8to16:Wminus1})\\
$W_{-2}$&word vector&third-last byte of each sequence in~$w_{\mathrm{in}}$&(\ref{8to16:Wminus2})\\
$W_{\mathrm{sum}}$&word vector&$W_{\mathrm{end}}$, $W_{-1}$, and~$W_{-2}$ bits assembled&(\ref{8to16:Wsum})\\
$W_{\mathrm{out}}$&word vector&$W_{\mathrm{end}}$ with surrogates fixed up&(\ref{8to16:Wout})\\
$M_{\mathrm{out}}$&word mask&valid words in~$W_{\mathrm{out}}$&(\ref{8to16:Mout})\\
$m_{\mathrm{processed}}$&byte mask&last byte of each sequence in~$M_{\mathrm{out}}$&(\ref{8to16:mprocessed})\\
$n_{\mathrm{in}}$&integer&number of $w_{\mathrm{in}}$~bytes processed&(\ref{8to16:nin})\\
$n_{\mathrm{out}}$&integer&number of words written out&(\ref{8to16:nout})\\
$\ell$&integer&number of input bytes left to process&---\\
$b$&byte mask&input bytes left to process&(\ref{8to16:b})\\
$m_c$&byte mask&where continuation bytes should be in~$w_{\mathrm{in}}$&(\ref{8to16:mc})\\
$m_{\mathrm{pre}}$&byte mask&$w_{\mathrm{in}}$ bytes preceding first mismatch&(\ref{8to16:mpre})\\
$M_{<\codepoint{800}}$&word mask&overlong 3-byte characters in~$W_{\mathrm{out}}$&(\ref{8to16:M800})\\
$M_{3s}$&word mask&3-byte characters encoding surr. in~$W_{\mathrm{out}}$&(\ref{8to16:M3s})\\
$M_{4s}$&word mask&surrogates not encoding surr. in~$W_{\mathrm{out}}$&(\ref{8to16:M4s})\\
\bottomrule
\end{tabular}
\end{equation*}

\noindent
\emph{Variables pertaining to the fast paths are not listed.}

\section{UTF-16 to UTF-8: Summary of Variables}
\label{appendix:utf16utf8}
\begin{equation*}
\begin{tabular}{ccp{0.54\textwidth}c}
\toprule
\emph{symbol}&\emph{type}&\emph{description}&\emph{Eq.}\\
\midrule
$W_{\mathrm{in}}$&word vector&input vector&---\\
$L$&word mask&position of lookahead word&(\ref{16to8:L})\\
$M_{234}$&word mask&words that are 2--4-byte characters&(\ref{16to8:M234})\\
$M_{12}$&word mask&words that are 1- and 2-byte characters&(\ref{16to8:M12})\\
$M_{\mathrm{hi}}$&word mask&words that are high surrogates&(\ref{16to8:Mhi})\\
$M_{\mathrm{lo}}$&word mask&words that are low surrogates&(\ref{16to8:Mlo})\\
$W_{\mathrm{lo}}$&dword vector&$W_{\mathrm{in}}$ shifted to the right by one element&---\\
$W_{\mathrm{joined}}$&dword vector&$W_{\mathrm{in}}$ with high and low surrogates joined&(\ref{16to8:Wjoined})\\
$W_{\mathrm{shifted}}$&dword vector&$W_{\mathrm{joined}}$ bits shifted with \texttt{vpmultishiftqb}&(\ref{16to8:Wshifted})\\
$W_{\mathrm{tag}}$&dword vector&UTF-8 tag bits for $W_{\mathrm{out}}$&(\ref{16to8:Wtag})\\
$W_{\mathrm{out}}$&dword vector&$W_{\mathrm{in}}$ transcoded to UTF-8 with padding&(\ref{16to8:Wout})\\
$W_{\mathrm{keep}}$&dword vector&magic constant for which bytes to keep&(\ref{16to8:Wkeep})\\
$m_{\mathrm{keep}}$&byte mask&mask of $W_{\mathrm{out}}$ bytes we want to keep&(\ref{16to8:mkeep})\\
$w_{\mathrm{out}}$&byte vector&output string without padding&(\ref{16to8:wout})\\
$n_{\mathrm{out}}$&integer&length of $w_{\mathrm{out}}$&(\ref{16to8:nout})\\
$c$&integer&surrogate carry (in)&---\\
$c_{\mathrm{out}}$&integer&surrogate carry out&(\ref{16to8:cout})\\

$M_{\mathrm{hi-lo}}$&word mask&high surrogates not followed by low surr.&(\ref{16to8:Mhilo})\\
$M_{\mathrm{lo-hi}}$&word mask&low surrogates not preceded by high surr.&(\ref{16to8:Mlohi})\\
$\ell$&integer&number of input words left to process&(\ref{16to8:l})\\
$B$&word mask&input words left to process&(\ref{16to8:B})\\
\bottomrule
\end{tabular}
\end{equation*}

\end{document}